# COSMOLOGICAL CONSTRAINTS FROM CLUSTER X–RAY MORPHOLOGIES

Joseph J. Mohr[†], August E. Evrard[‡], Daniel G. Fabricant[†], & Margaret J. Geller[†]

## ABSTRACT

We use a representative sample of 65 galaxy clusters observed with the *Einstein* IPC to constrain the range of cluster X–ray morphologies. We develop and apply quantitative and reproducible measures to constrain the intrinsic distributions of (i) emission weighted centroid variation $w_{\vec{x}}$, (ii) emission weighted axial ratio $\eta$, (iii) emission weighted orientation $\theta_o$, and (iv) measures of the radial fall–off, $\alpha$ and $\beta$. For each cluster we use a Monte–Carlo procedure to determine the effects of Poisson noise, detector imperfections, and foreground/background X–ray point sources.

We then use the range of cluster X–ray morphologies to constrain three generic cosmological models ($\Omega$=1, $\Omega_o$=0.2, and $\Omega_o$=0.2 & $\lambda_o$=0.8). For each of these models, we evolve eight sets of Gaussian random initial conditions consistent with an effective power spectrum $P(k) \propto k^{-1}$ on cluster scales. Using this sample of 24 numerical cluster simulations which include gravity and gas physics (but no cooling or ejection from galaxies), we compare the X–ray morphologies of the observed and simulated clusters. Specifically, we build artificial ensembles with the same distributions in the number of cluster photons, X–ray temperature, and cluster redshift as the *Einstein* ensemble; we then compare the observed and simulated distributions in $w_{\vec{x}}$, $\eta$, and $\alpha$.

The comparisons indicate that: (i) these three morphological characteristics are sensitive to the underlying cosmological model, and (ii) galaxy clusters with the observed range of X–ray morphologies are very unlikely in low $\Omega_0$ cosmologies. The analysis favors the $\Omega$=1 model, though some discrepancies remain. We discuss the effects of changing the initial conditions and of including additional physics in the simulations.

## 1. INTRODUCTION

Over the past decade, studies have provided ample evidence that a significant fraction of galaxy clusters have undergone recent growth (Geller & Beers 1982, Dressler & Shectman 1988, Forman & Jones 1990). Analytic work by Richstone, Loeb, & Turner (1992) demonstrates that the present epoch dynamical state of galaxy clusters is sensitive to the cosmological density parameter $\Omega_0$. We have previously shown that the structure of the

† Harvard–Smithsonian Center for Astrophysics, 60 Garden St., Cambridge, MA 02138
‡ Physics Department, University of Michigan, 1049 Randall Lab, Ann Arbor, MI 48109

X–ray emitting gas in clusters reflects this dependence (Evrard *et al.* 1993). Thus cluster X–ray morphologies provide an important cosmological constraint.

Here we use *quantitative* and *reproducible* measures of cluster X–ray morphology for a representative sample of galaxy clusters to evaluate these cosmological constraints. We first develop and test measures of the cluster centroid variation $w_{\vec{x}}$ (Mohr, Fabricant, & Geller 1993; hereafter MFG93), the axial ratio $\eta$, the orientation $\theta_o$, and the measures of radial fall–off, $\alpha$ and $\beta$. We apply these measures to a representative sample of cluster X–ray observations to determine the intrinsic cluster distribution in these quantities (§2). Finally, we directly compare the morphologies of observed clusters to a set of clusters simulated within three different cosmological models; we use the results to place cosmological constraints (§3). §4 contains a summary and further discussion of the results. (A Hubble parameter of $H_0 = 50\ h_{50}^{-1}$ km s$^{-1}$ Mpc$^{-1}$ is assumed throughout.)

## 2. THE MORPHOLOGIES OF OBSERVED CLUSTERS

This section is a description of the measures of cluster X–ray morphology. We include a discussion of the observational sample taken from the *Einstein* archive, and a brief review of the standard IPC reduction procedure. Case by case, we provide the measurement details of four morphological characteristics: (i) the centroid variation $w_{\vec{x}}$, (ii) the axial ratio $\eta$, (iii) the orientation $\theta_o$, and (iv) two measures of the radial fall–off, $\alpha$ and $\beta$. We review the tests of our measurement accuracy and display the results.

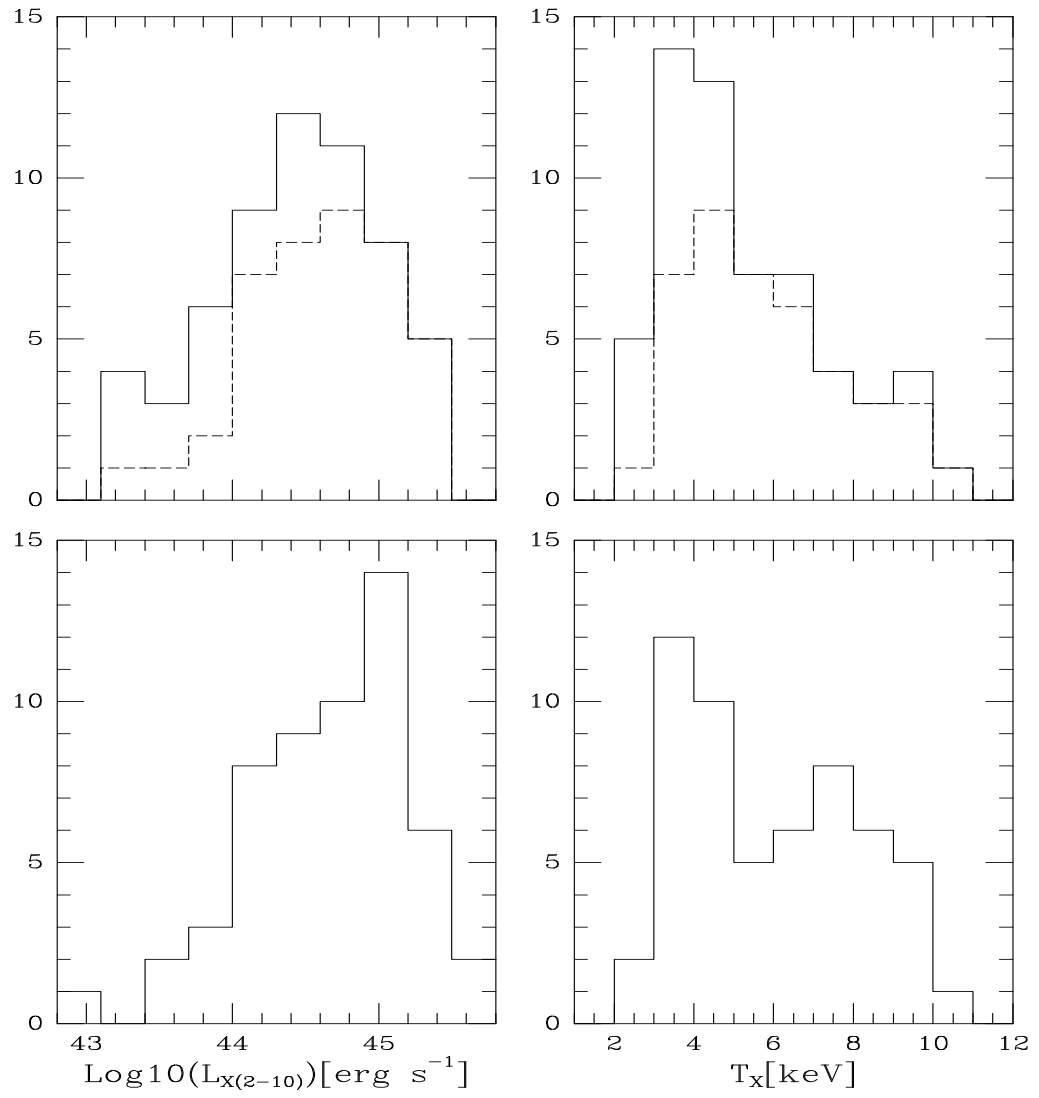

Figure 1: Completeness of the *Einstein* sample. Column 1 contains distributions of 2–10 keV X–ray luminosity for (bottom) the 55 members of the flux limited Edge *et al.* (1990) sample and (top) the 58 *Einstein* clusters with published luminosities (Edge *et al.* 1990, David *et al.* 1992). The dashed lines in the upper plot form the distribution for the 41 of 58 *Einstein* clusters contained in the Edge *et al.* flux limited sample. Column 2 contains an analogous arrangement of the distributions in X–ray temperature. The KS test demonstrates that there is a reasonable probability that these three samples are drawn from the same parent distribution.

### 2.1 THE CLUSTER SAMPLE

Cluster X–ray morphologies are quite varied. To probe the full range of morphologies it is imperative to have a large, representative sample of high signal–to–noise cluster images. In the near future the ROSAT public archive will be a valuable resource in this regard. For the present study we extract and reduce *Einstein* Imaging Proportional Counter (IPC) images of 65 galaxy clusters. We choose these particular clusters in an attempt to construct a representative sample and because they are among the best cluster images obtained with the IPC. Our ensemble of observations contains images with from $10^3$ to $2 \times 10^5$ cluster photons; the median number of cluster photons per image is $\sim$5000. Because this cluster sample is not manifestly complete with respect to volume or flux limits, we evaluate the effects of incompleteness by comparing the sample to the X–ray flux limited sample of 55 clusters constructed by Edge *et al.* (1990).

Of our 65 clusters, 58 have published 2–10 keV X–ray luminosities ($L_X$) and X–ray temperatures ($T_X$) (Edge *et al.* 1990; David *et al.* 1992). 41 of these 58 clusters are in the Edge *et al.* sample. Figure 1 compares the distributions in $L_X$ (left column) and $T_X$ (right column) of our ensemble (top) with the Edge sample (bottom). A first glance at the distributions may lead to the conclusion that the entire *Einstein* sample has more low luminosity and low temperature clusters than both the Edge sample and the *Einstein* subsample. However, the KS test demonstrates that the differences in these distributions are all consistent with statistical fluctuations. For each combination we include the KS parameter D and the implied probability that the two distributions are consistent. For $T_X$: Edge/Ein58 D=0.16 (45%), Edge/Ein41 D=0.10 (98%), Ein41/Ein58 D=0.15 (69%). For $L_X$: Edge/Ein58 D=0.21 (15%), Edge/Ein41 D=0.14 (71%), Ein41/Ein58 D=0.15 (67%). On the basis of these results, the $L_X$ and $T_X$ distributions for both the entire *Einstein* sample of 58 clusters and the *Einstein* subsample of 41 clusters are indistinguishable from an X–ray flux limited sample.

Naturally, we would like to have images of comparable quality and depth for all 65 clusters. However, variations in exposure time and surface brightness within our sample complicate our goal of extracting the intrinsic range of cluster morphologies. Using morphological measures which are most sensitive to the bright cluster cores reduces sensitivity to these sample variations. Each of our measures (except for the radial fall–off) is emission weighted and therefore most sensitive to the highest signal–to–noise regions of the X–ray images. Our measures are relatively insensitive to the inclusion or exclusion of outer cluster regions. Indeed, for a cluster with the mean radial behavior ($\beta = 0.65$), the emission weighting scheme outside the cluster core is essentially a radial weighting by $R^{-2}$ where $R$ is the projected distance from the peak in the X–ray surface brightness.

## 2.2 IMAGE REDUCTION

We filter the observation to produce a 0.3 to 3.5 keV X–ray image blocked from the original $8''$pixels to $16''$pixels. We then subtract an approximation to the image background using a scaled Deep Survey background map; the background is scaled using either the observation livetime or the X–ray flux in regions of the image outside the cluster. Then we apply corrections for telescope vignetting and small scale IPC non–linearities (see Soltan & Fabricant (1990) and MFG93 for a discussion of the small scale correction). Finally, we block the image to $32''$pixels and smooth with a Gaussian of $1.5'$ FWHM. The resulting image has a spatial resolution of $2.4'$ FWHM. Our analysis of these images is confined to the detector region within the IPC ribs.

## 2.3 CENTROID VARIATION

We analyze each X–ray image in annuli of constant width ($1.07'$) and increasing radius; as described below, these annuli are not concentric. Because an *a priori* choice of image center biases the results and the cluster centroid often varies as a function of scale, the first step in the analysis is defining a cluster center for each annulus. We use a simplex minimization routine (Press *et al.* 1988) to find the annulus position which minimizes the distance between the annulus centroid $\vec{x}$ and the geometric center of the annulus. $\vec{x}$ is the intensity weighted first moment of the annulus

$$\vec{x} = \left( \frac{1}{N} \sum_i n_i x_i, \; \frac{1}{N} \sum_i n_i y_i \right) \quad \text{where} \quad N = \sum_i n_i, \tag{2.1}$$

$n_i$ is the number of photons in pixel $i$ and the summation occurs over all pixels within the annulus. This minimization is roughly analogous to balancing a donut on an anthill– the final position of the donut is the position which brings the plane defined by the donut as nearly parallel to the ground as the anthill allows (see MFG93 for more details). Once the appropriate annulus position is determined, we use the centroid $\vec{x}$ calculated at that position as the image centroid for that annulus radius.

The annulus radius is increased and the identical prescription followed until measurements cover the useful portion of the X–ray image (determined by the IPC ribs and any bright point sources). Thus we extract a radial function in the centroid $\vec{x}$. To examine the variation in this function while minimizing the noise introduced by the low signal–to–noise region of the image, we calculate an emission weighted centroid variation over the region of the image which satisfies a mean signal–to–noise constraint ($\langle \frac{S}{N} \rangle > 5$ where $\langle \frac{S}{N} \rangle$ is the mean signal to noise of all the pixels within the annulus). Specifically, the emission weighted centroid variation is

$$w_{\vec{x}}^2 = \left( \sum_j N_j \right)^{-1} \sum_j N_j \left( \vec{x}_j - \langle \vec{x} \rangle \right)^2 \tag{2.2}$$

where the summation occurs over all annuli with $\langle \frac{S}{N} \rangle > 5$, $N_j$ is the number of photons within the annulus $j$, and $\langle \vec{x} \rangle$ is the emission weighted average centroid of the image.

We take a Monte–Carlo approach to determine the significance of the centroid variation. We ask, "Is the centroid variation we detect consistent with variations introduced by Poisson noise, foreground/background point sources and detector imperfections?" To answer this question, we first fit our cluster image to a symmetric $\beta$ model of the form

$$I(x,y) = I_0 \left( 1 + \left( \frac{x}{R_c} \right)^2 + \left( \frac{y}{\eta R_c} \right)^2 \right)^{\frac{1}{2} - 3\beta} \tag{2.3}$$

where $I_0$, $R_c$, $\eta$ and $\beta$ are the normalization, core radius, axial ratio and radial fall–off. Two other free parameters, a cluster orientation angle and emission center on the IPC, are included because the detector response varies over its face. This $\beta$ model has no intrinsic centroid variation.

We then simulate an observation of this model, including X–ray background, Poisson noise, detector response variations and randomly positioned point sources consistent with the EMSS log $N$–log $S$ distribution (Gioia *et al.* 1990; see MFG93 for details). We reduce the simulated image ("$\beta$–image" hereafter) just as we reduce the original image and carry out the measurements described above. By reimaging the $\beta$ model from 200 to 1000 times we build up a sample of $\beta$–images with a range of morphologies (and centroid variations) due solely to non–cluster effects. We use this distribution of centroid variations in two different ways: (i) we use the mean of the distribution $\langle w_{\vec{x}}^2 \rangle_{MC}$ to correct the centroid variation measured in the original image for the Poisson, point source and instrumental contributions ($w_{\vec{x}}^2 = w_{meas}^2 - \langle w_{\vec{x}}^2 \rangle_{MC}$ where $w_{meas}^2$ and $w_{\vec{x}}^2$ are the original and corrected centroid variations) and (ii) we use the width of the Monte–Carlo distribution as one component of the uncertainty in the centroid variation.

The value of $w_{\vec{x}}$ depends on the region of the cluster used to calculate it (as discussed later, the orientation, axial ratio, and radial fall–off share this characteristic). This dependence can be produced by a cluster cooling flow located off center with respect to the large scale structure of the cluster. In this case the cluster centroid would vary significantly near the cluster core and then vary little outside the core. Thus, $w_{\vec{x}}$ would decrease as we included more and more annuli from the outer regions of the cluster. Because the centroid variation depends on the radial scale of the measurement, the choice of an appropriate scale is crucial. Here we interpret a centroid variation as an indicator of dynamical youth (MFG93, Mohr & Evrard, 1994). Thus, we take the radius where $w_{\vec{x}}$ reaches maximum significance ($max\{w_{\vec{x}}/\sigma_w\}$, where $\sigma_w$ is the centroid variation uncertainty).

In §3 we demonstrate that clusters simulated with an Einstein–deSitter cosmological model have morphologies very similar to observed clusters. Thus these cluster simulations provide an ideal medium to test our measurement accuracy. We compare measurements made on perfect images of the simulated clusters (no instrumental effects or Poisson noise) with the measurements made on artificial IPC images of these clusters (see §3 for details about these artificial images). In particular, we use a sample of more than 200 cluster images with distributions in signal–to–noise and spatial scale similar to those of the *Einstein* IPC sample.

Figure 2 contains the results of these comparisons. The first column contains a measure of the accuracy of $w_{\vec{x}}$; the top row contains a plot of the fractional error as a function of the real value. For a centroid variation $w_{\vec{x}} < 0.3'$, the accuracy decreases because of the resolution and noise characteristics of our image ensemble (and because we plot the fractional error which is proportional to ${w_{\vec{x}}}^{-1}$). The large number of photons in a typical image (~5000) allows us to extract variations which are $\sim \frac{1}{3}$ the resolution scale of the images. Greater resolution would require either IPC images with significantly more cluster photons or an instrument with improved imaging and background characteristics (e.g. the ROSAT PSPC). The bias for images with $w_{real} > 0.3'$ is $\langle\Delta w/w_{real}\rangle = -7.6\%$; on average we tend to underestimate the magnitude of the centroid variation. This bias results from our method of correcting for the Poisson and instrumental contributions to $w_{meas}$. As described above, we postulate that the measured centroid variation has an intrinsic and a noise component: $w_{meas}^2 = w_{\vec{x}}^2 + \langle w_{\vec{x}}^2\rangle_{MC}$. This approximation works well at large centroid variation but leads to an underestimate for small variations. Clearly, Poisson noise and instrumental effects can also *mask* a cluster centroid variation!

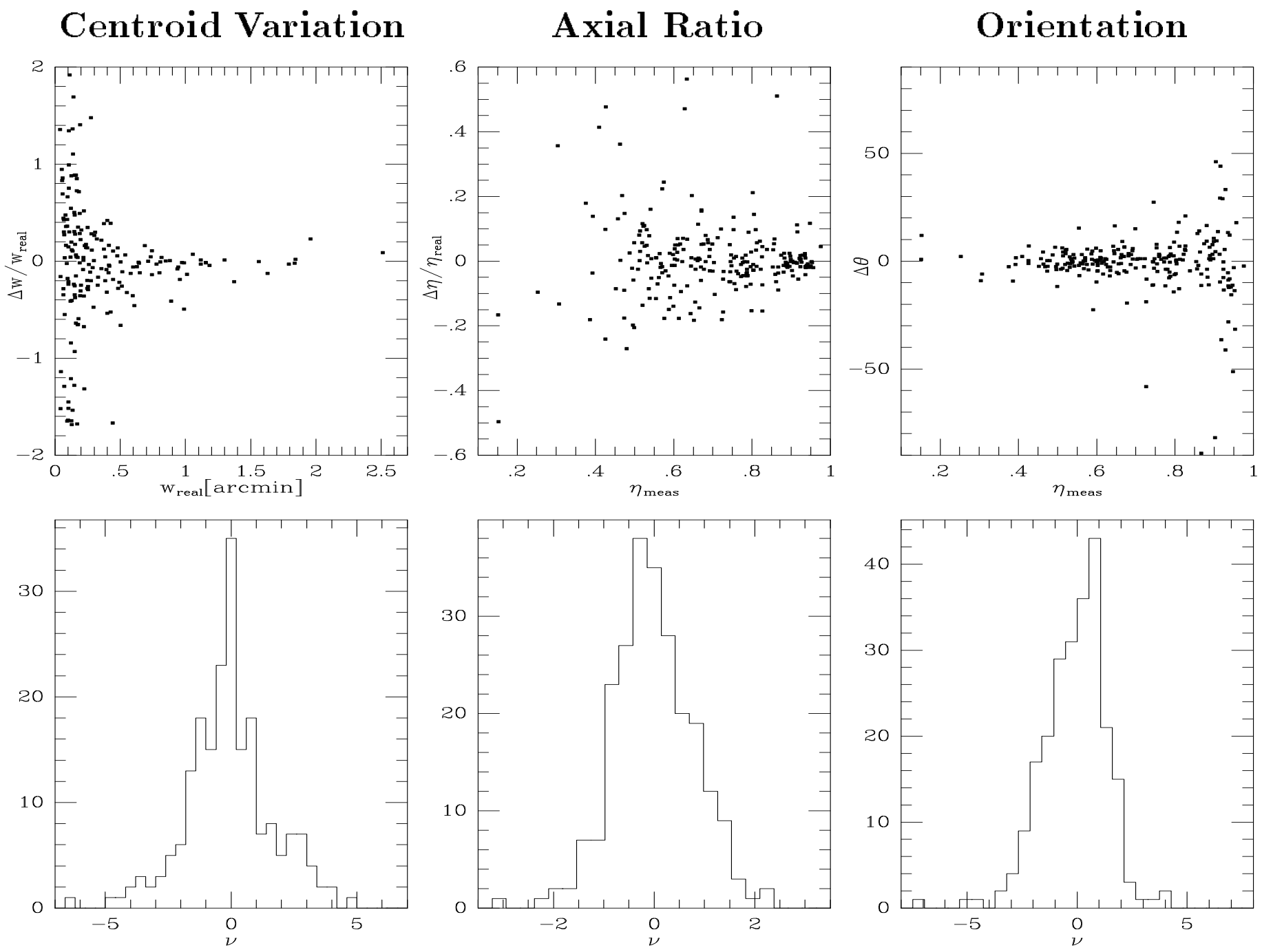

Figure 2: Measurement accuracy evaluated using a set of over 200 artificial IPC images. The first row contains (from left to right) (i) the fractional error in the centroid variation ($\Delta w/w_{real} = (w_{\vec{x}} - w_{real})/w_{real}$) versus the real centroid variation $w_{real}$, (ii) the fractional error in the axial ratio ($\Delta\eta/\eta_{real} = (\eta_{meas} - \eta_{real})/\eta_{real}$) versus the measured axial ratio $\eta_{meas}$, and (iii) the absolute error in the orientation angle ($\Delta\theta$ in degrees) versus the measured axial ratio $\eta_{meas}$. The mean errors for the three parameters are $\langle\Delta w/w_{real}\rangle = -7.6\%$ (for all $w_{real} > 0.3'$), $\langle\Delta\eta/\eta_{real}\rangle = -0.7\%$, and $\langle\Delta\theta\rangle = -0.85°$. The second row contains the histogram of scaled errors ($\nu = \Delta/\sigma$, where $\Delta$ is the error and $\sigma$ is the uncertainty) for each measured parameter. The best Gaussian fits to the $\nu$ distributions have the widths $\sigma_w = 1.34$, $\sigma_\eta = 0.73$, and $\sigma_\theta = 1.49$. For each parameter we use these widths to correct our uncertainties in order to obtain a $\nu$ distribution of nominal width.

The bottom row of the first column of Figure 2 contains a scaled error distribution for all images ($\nu = \Delta/\sigma$ where $\Delta$ is the measurement error and $\sigma$ is the measurement uncertainty). We obtain this distribution by including three contributions in the centroid variation uncertainty: (i) the width of the Monte–Carlo distribution of $w_{\vec{x}}$, (ii) the formal uncertainties which follow directly from Equation 2.1 and Equation 2.2, and (iii) a 10% fractional uncertainty which accounts for the uncertainties in the IPC response. This flat 10% fractional uncertainty removes the outliers in our scaled error distribution. A Gaussian fit to the $\nu$ distribution demonstrates that we underestimate the uncertainties by 34% (the best fit $\sigma = 1.34$). We increase all $w_{\vec{x}}$ uncertainties by 34% to correct for this underestimate.

Unfortunately, it is not possible to extract a centroid variation from all 65 cluster images. Table 1 lists $w_{\vec{x}}$ for 46 clusters and the maximum centroid shift $\Delta\vec{x}$ within the $\left\langle\frac{S}{N}\right\rangle > 5$ region for 58 clusters. The table contains the centroid variation and uncertainty in arcmin ($w_{\vec{x}}[']$ and $\sigma_w$), the centroid variation and uncertainty in $h_{50}^{-1}$kpc ($w_{\vec{x}}[kpc]$ and $\sigma_w$), the radius (in arcmin) of the maximally significant $w_{\vec{x}}$ measurement ($R[']$), and the maximum centroid shift ($\Delta\vec{x}$). Clusters with a "-" entry have centroid variations consistent with zero (where $\langle w_{\vec{x}}^2 \rangle_{MC}$ is greater than $w_{meas}^2$). A "*" marks those cluster where the $\langle \frac{S}{N} \rangle > 5$ region is too small to make a meaningful $w_{\vec{x}}$ measurement, and a "p" marks clusters with bright point sources near their emission peaks, making $w_{\vec{x}}$ difficult to measure with the *Einstein* IPC. (The higher resolution ROSAT PSPC images of these clusters may allow accurate measurements after point source removal.) 43 of these 46 clusters have measured $L_X$ and $T_X$. A KS test demonstrates that this sample of 43 clusters is statistically indistinguishable from the flux limited sample of Edge *et al.* ($T_X$: D=0.20 (28%), $L_X$: D=0.23 (15%)). Figure 3 shows the distribution of $w_{\vec{x}}$. The mean and root mean square width of this distribution are $\langle w_{\vec{x}} \rangle = 41.3$ kpc and RMS$_w$ = 47.4 kpc. Notes on individual measurements are in Appendix A.

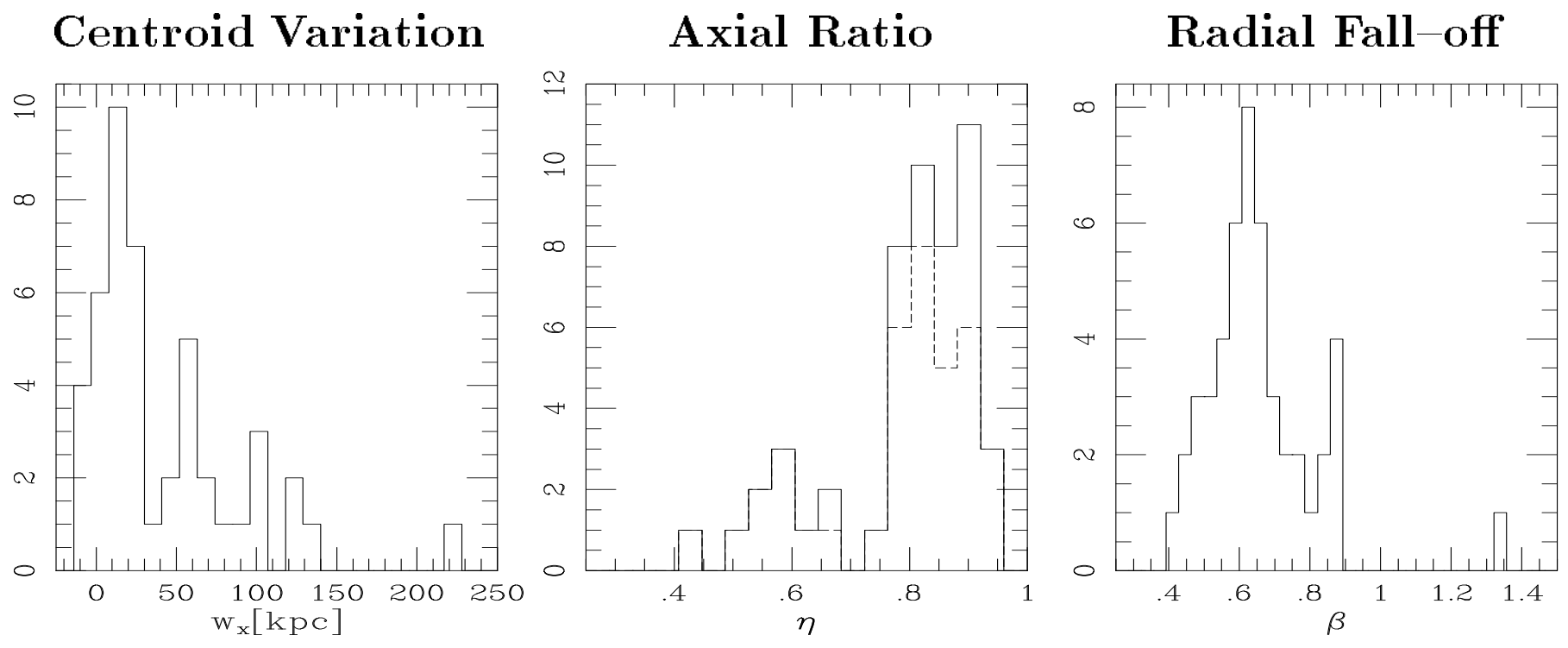

Figure 3: X–ray morphologies of observed clusters. From left to right is (i) the distribution of emission weighted centroid variation $w_{\vec{x}}$ in kpc for a sample of 43 clusters, (ii) the distribution of emission weighted axial ratios $\eta$ for a sample of 51 clusters and (iii) the distribution of radial fall–off $\beta$ for 48 clusters. The mean and root mean square width of the three distributions are: a) $\langle w_{\vec{x}} \rangle = 41.3$ kpc and RMS$_w$ = 47.4 kpc, b) $\langle \eta \rangle = 0.80$ and RMS$_\eta$ = 0.12, and c) $\langle \beta \rangle = 0.65$ and RMS$_\beta$ = 0.16. The dashed line histogram in the $\eta$ distribution contains 38 clusters. The 13 clusters removed from the original sample have central cooling times significantly less than $10^{10}$ years (Edge, Stewart, & Fabian 1992, Stewart *et al.* 1984).

Because the centroid variation is a signature of dynamical youth, we can use these measurements to examine the fraction of substructure in galaxy cluster cores. Of the 46 clusters with measured $w_{\vec{x}}$, 71% have centroid variations which are at least $3\sigma_w$ from zero; 48% have centroids variations which are at least $5\sigma_w$ from zero. The Monte–Carlo method we use to determine the uncertainties in our measurements also gives us the probability that the measured centroid variation in a cluster is merely due to Poisson noise, detector imperfections, and X–ray point sources. The probability that $w_{\vec{x}}$ is due to an intrinsic, cluster centroid variation is $> 99.9\%$ in 46% of the sample and is $> 99\%$ in 61% of the sample. Thus, we conclude that the IPC images of these 46 clusters provide evidence for core substructure in from 50% to 70% of the sample, depending on the confidence criterion adopted.

We include $\Delta\vec{x}$ because this scale is readily apparent in a contour plot of the cluster X–ray emission. Table 1 lists the values for all clusters with at least two meaningful centroid measurements within the $\langle \frac{S}{N} \rangle > 5$ region of the image. We avoid using $\Delta\vec{x}$ for comparison of X–ray morphologies because it is a noisier measure than $w_{\vec{x}}$ (dependent on only two centroid

Table 1. Centroid Variation

| Cluster | $w_{\vec{x}}[']$ | $\sigma_w$ | $R[']$ | $w_{\vec{x}}$[kpc] | $\sigma_w$ | $\Delta\vec{x}[']$ | Cluster | $w_{\vec{x}}[']$ | $\sigma_w$ | $R[']$ | $w_{\vec{x}}$[kpc] | $\sigma_w$ | $\Delta\vec{x}[']$ |
|---|---|---|---|---|---|---|---|---|---|---|---|---|---|
| A85$^{ec}$ | 0.504 | 0.079 | 10.7 | 45.8 | 7.2 | 1.53 | A2065$^{e}$ | 0.159 | 0.063 | 7.5 | 20.0 | 7.9 | 0.55 |
| A119$^{e}$ | 0.227 | 0.173 | 6.4 | 17.6 | 13.4 | 0.92 | A2124$^{m}$ | * | * | | | | 0.34 |
| A133 | 0.226 | 0.058 | 6.4 | 23.8 | 6.0 | 0.72 | A2142$^{ec}$ | p | p | | | | |
| A168 | 1.254 | 0.230 | 8.5 | 98.0 | 18.0 | 3.82 | A2147$^{e}$ | 1.565 | 0.236 | 12.8 | 96.4 | 14.5 | 4.32 |
| A262$^{ec}$ | 0.127 | 0.059 | 6.4 | 3.6 | 1.6 | 0.39 | A2151 | 0.914 | 0.135 | 7.5 | 59.0 | 8.7 | 2.50 |
| A399$^{e}$ | 0.610 | 0.109 | 8.5 | 76.9 | 13.5 | 1.91 | A2199$^{ec}$ | 0.141 | 0.025 | 6.4 | 7.4 | 1.3 | 0.48 |
| A400 | 0.405 | 0.154 | 8.5 | 16.8 | 6.4 | 1.39 | A2255$^{e}$ | 0.174 | 0.064 | 6.4 | 24.6 | 9.1 | 0.74 |
| A401$^{e}$ | 0.492 | 0.072 | 9.6 | 63.5 | 9.4 | 1.61 | A2256$^{e}$ | 0.535 | 0.084 | 12.8 | 54.2 | 8.6 | 1.66 |
| A426$^{ec}$ | 0.915 | 0.125 | 13.9 | 28.6 | 3.9 | 2.23 | A2319$^{e}$ | 0.524 | 0.072 | 7.5 | 50.7 | 7.1 | 1.93 |
| A478$^{ec}$ | - | - | 6.4 | | | 0.19 | A2410$^{m}$ | 0.631 | 0.860 | 6.4 | 88.2 | 120.3 | 2.59 |
| A496$^{ec}$ | 0.392 | 0.066 | 9.6 | 22.6 | 3.8 | 1.03 | A2420 | 0.197 | 0.059 | 7.5 | 28.8 | 8.6 | 0.77 |
| A539 | p | p | | | | | A2440 | * | * | | | | 0.12 |
| A548 | 0.221 | 0.090 | 6.4 | 16.0 | 6.6 | 0.69 | A2593 | * | * | | | | 0.98 |
| A576$^{e}$ | 0.180 | 0.058 | 7.5 | 12.0 | 3.8 | 0.55 | A2597$^{e}$ | * | * | | | | 0.14 |
| A592$^{m}$ | * | * | | | | 0.38 | A2626 | * | * | | | | |
| A644$^{e}$ | 0.464 | 0.079 | 8.5 | 57.6 | 9.9 | 1.33 | A2634 | - | - | 8.5 | | | 3.40 |
| A671$^{m}$ | 0.145 | 0.044 | 6.4 | 12.7 | 3.9 | 0.47 | A2657$^{c}$ | 0.193 | 0.048 | 6.4 | 13.5 | 3.4 | 1.08 |
| A754$^{e}$ | 1.082 | 0.153 | 9.6 | 102.3 | 14.3 | 3.50 | A2670 | * | * | | | | 0.24 |
| A780$^{ec}$ | - | - | 8.5 | | | 0.32 | A2877$^{c}$ | p | p | | | | |
| A1060$^{e}$ | 0.267 | 0.046 | 8.5 | 5.8 | 0.9 | 0.78 | A3158$^{e}$ | 0.376 | 0.063 | 8.5 | 38.7 | 6.6 | 1.24 |
| A1367$^{e}$ | 0.832 | 0.170 | 12.8 | 27.6 | 5.6 | 3.39 | A3186$^{m}$ | * | * | | | | |
| A1644$^{e}$ | 1.527 | 0.208 | 9.6 | 126.5 | 17.2 | 4.20 | A3266$^{e}$ | 0.574 | 0.083 | 11.7 | 54.5 | 7.9 | 1.34 |
| A1650$^{e}$ | * | * | | | | 0.24 | A3376$^{m}$ | 0.929 | 0.307 | 9.6 | 73.8 | 24.3 | 3.08 |
| A1656$^{e}$ | 0.429 | 0.066 | 14.9 | 17.3 | 2.7 | 1.25 | A3391$^{e}$ | - | - | 8.5 | | | 0.71 |
| A1689$^{e}$ | * | * | | | | 0.07 | A3395SW | 2.531 | 0.341 | 6.4 | 225.1 | 30.3 | 5.96 |
| A1736$^{e}$ | 0.771 | 0.118 | 9.6 | 62.0 | 9.5 | 2.53 | A3526$^{ec}$ | 0.451 | 0.079 | 9.6 | 7.9 | 1.3 | 1.24 |
| A1767 | - | - | 6.4 | | | 0.46 | A3532$^{e}$ | * | * | | | | 0.66 |
| A1775 | * | * | | | | 0.16 | A3667$^{e}$ | 1.371 | 0.186 | 7.5 | 126.7 | 17.2 | 1.40 |
| A1795$^{ec}$ | p | p | | | | | AWM7$^{ec}$ | 0.320 | 0.087 | 6.4 | 9.6 | 2.5 | 0.93 |
| A1983$^{m}$ | * | * | | | | 0.23 | CygA$^{ec}$ | 1.405 | 0.193 | 9.6 | 139.7 | 19.2 | 4.37 |
| A2029$^{ec}$ | p | p | | | | | MKW3S$^{ec}$ | * | * | | | | 0.07 |
| A2052$^{ec}$ | - | - | 6.4 | | | 0.20 | 3C129$^{e}$ | 0.413 | 0.121 | 10.7 | 15.8 | 4.7 | 1.39 |
| A2063$^{ec}$ | 0.230 | 0.063 | 7.5 | 14.2 | 3.9 | 0.72 | | | | | | | |

- $w_{\vec{x}}$ consistent with zero
$*$ $\left\langle \frac{S}{N} \right\rangle > 5$ region too small
p bright point source near core

$^{e}$ member of Edge *et al.* sample
$^{m}$ no measured X–ray temperature
$^{c}$ central cooling time $< 10^{10}$years

measurements rather than an ensemble of >4 measurements).

## 2.4 AXIAL RATIO

The methods of measuring the cluster axial ratio and the centroid variation are similar. Given the correct position for an annulus, we expand the photon distribution defined by the annulus in a finite Fourier series.

$$I(\theta) = \sum_{n=0}^{4} A_n \cos n\theta + B_n \sin n\theta \qquad (2.4)$$

The series is truncated at $n = 4$ to reach a balance between execution speed and accuracy in the low $n$ terms. The axial ratio comes from the coefficients of the $2\theta$ terms:

$$\eta_{\bar{r}} = \frac{r_{minor}}{r_{major}} = \frac{\bar{r} - \delta r}{\bar{r} + \delta r} \approx 1 - \frac{2\sqrt{A_2^2 + B_2^2}}{r\frac{dI}{dr}} \tag{2.5}$$

A formal uncertainty for the axial ratio follows from its dependence on $A_2$ and $B_2$ and the appropriate diagonal elements of the covariance matrix from the expansion. The final uncertainty for the axial ratio at a particular radius is the sum in quadrature of the formal uncertainty and a more realistic uncertainty based on the Monte–Carlo approach (§2.3). The Monte–Carlo uncertainty is the RMS variation of the axial ratio over the ensemble of $\beta$–images.

With this approach we measure the cluster axial ratio (and uncertainty) as a function of radius. We use this function within the $\langle \frac{S}{N} \rangle > 5$ region of the cluster image to calculate $\eta$, the emission weighted average axial ratio.

$$\eta = \left[ \sum_j \left( \frac{N}{\sigma_\eta^2} \right)_j \right]^{-1} \sum_j \left( \frac{N\eta_{\bar{r}}}{\sigma_\eta^2} \right)_j \tag{2.6}$$

where $N/\sigma_\eta^2$ is the number of photons scaled by the axial ratio uncertainty in each annulus and $\eta_{\bar{r}}$ is the measured axial ratio in each annulus. As for $w_{\vec{x}}$, we use a large ensemble of numerical cluster simulations to evaluate the accuracy of our measurements. The second column of Figure 2 contains the results for $\eta$. The top row contains a plot of the fractional error in $\eta$ as a function of the measured axial ratio. There is no significant bias in this measurement; the mean error in the more than 200 images is $\langle \Delta\eta/\eta_{real} \rangle = -0.2\%$. The width of the best fit Gaussian to the scaled error distribution in the bottom row is $\sigma$=0.73, 27% less than the nominal width. We reduce the uncertainties by 27% to correct for this overestimate.

Columns labeled $\eta$ and $\sigma_\eta$ of Table 2 contain the list of axial ratios and uncertainties for 57 clusters. The columns labeled $R[']$ contain the radius (in arcmin) within which we calculate the emission weighted average. 51 of these clusters have measured $L_X$ and $T_X$ which we use to compare this sample to the flux limited sample of Edge *et al.* The KS test demonstrates that this sample of 51 clusters is statistically indistinguishable from the Edge *et al.* sample ($T_X$: D=0.19 (31%), $L_X$: D=0.24 (10%)). Figure 3 contains a histogram of $\eta$ for those 51 clusters; the mean and width of this distribution are $\langle\eta\rangle = 0.80$ and $\mathrm{RMS}_\eta = 0.12$. Appendix A contains a detailed discussion of individual measurements.

McMillan, Kowalski, & Ulmer (1989– hereafter MKU) study the orientation and axial ratio of the X–ray emission from 49 Abell clusters. They focus on the faint outer region of the cluster image. Specifically, the region they use is defined by two isophotes: a low flux level just above the noise and a high flux level defined so the entire region contains 20% of the cluster flux. Because we emphasize the core of the X–ray emission, our results differ from those of MKU in cases where the morphology of the X–ray emission varies with radius.

Table 2. Radial Fall–off, Axial Ratio and Orientation

| Cluster | $\beta^{\ddagger}$ | $\eta$ | $\sigma_\eta$ | $\theta_o$ | $\sigma_\theta$ | $R[']$ | Cluster | $\beta^{\ddagger}$ | $\eta$ | $\sigma_\eta$ | $\theta_o$ | $\sigma_\theta$ | $R[']$ |
|---|---|---|---|---|---|---|---|---|---|---|---|---|---|
| A85$^{ec}$ | 0.592 | 0.910 | 0.011 | -11 | 4 | 5.3 | A2065$^{e}$ | 0.767 | 0.764 | 0.031 | -27 | 2 | 8.5 |
| A119$^{e}$ | 0.735 | 0.791 | 0.066 | 2 | 8 | 10.7 | A2124$^{m}$ | 0.512 | 0.830 | 0.120 | -3 | 26 | 4.3 |
| A133 | 0.659 | 0.915 | 0.027 | -4 | 10 | 6.4 | A2142$^{ec}$ | 0.888 | | | | | |
| A168 | | 0.506 | 0.108 | -15 | 8 | 8.5 | A2147$^{e}$ | | 0.558 | 0.092 | -2 | 10 | 10.7 |
| A262$^{ec}$ | 0.521 | 0.835 | 0.041 | 39 | 5 | 9.6 | A2151 | | 0.910 | 0.042 | -56 | 7 | 6.4 |
| A399$^{e}$ | 0.578 | 0.854 | 0.039 | 16 | 5 | 8.5 | A2199$^{ec}$ | 0.594 | 0.823 | 0.019 | 35 | 2 | 11.7 |
| A400 | | 0.680 | 0.079 | 16 | 8 | 8.5 | A2255$^{e}$ | 0.838 | 0.855 | 0.036 | -80 | 13 | 9.6 |
| A401$^{e}$ | 0.627 | 0.798 | 0.022 | 23 | 4 | 9.6 | A2256$^{e}$ | | 0.738 | 0.019 | -64 | 2 | 12.8 |
| A426$^{ec}$ | 0.485 | 0.911 | 0.006 | 69 | 1 | 13.9 | A2319$^{e}$ | 0.622 | 0.828 | 0.022 | -27 | 2 | 12.8 |
| A478$^{ec}$ | 0.668 | 0.855 | 0.018 | 42 | 5 | 6.4 | A2410$^{m}$ | | | | 72 | 2 | 6.4 |
| A496$^{ec}$ | 0.641 | 0.877 | 0.018 | -11 | 4 | 10.7 | A2420 | 0.882 | 0.853 | 0.033 | 61 | 7 | 7.5 |
| A539 | 0.561 | | | | | | A2440 | | | | 37 | 10 | 4.3 |
| A548 | 0.489 | 0.886 | 0.054 | 20 | 10 | 7.5 | A2593 | 0.536 | 0.817 | 0.106 | -7 | 29 | 5.3 |
| A576$^{e}$ | 0.641 | 0.798 | 0.031 | 15 | 10 | 9.6 | A2597$^{e}$ | 0.870 | 0.934 | 0.029 | -51 | 14 | 4.3 |
| A592$^{m}$ | 0.569 | 0.862 | 0.050 | -78 | 13 | 5.3 | A2626 | 0.885 | 0.822 | 0.064 | 14 | 20 | 3.2 |
| A644$^{e}$ | 0.735 | 0.822 | 0.019 | 9 | 2 | 8.5 | A2634 | 0.651 | 0.528 | 0.131 | 36 | 7 | 8.5 |
| A671$^{m}$ | 0.743 | 0.917 | 0.029 | 36 | 23 | 6.4 | A2657$^{c}$ | 0.597 | 0.887 | 0.030 | 90 | 5 | 6.4 |
| A754$^{e}$ | | 0.579 | 0.042 | -87 | 2 | 10.7 | A2670 | 0.681 | 0.916 | 0.047 | -50 | 28 | 4.3 |
| A780$^{ec}$ | 0.681 | 0.915 | 0.014 | -22 | 4 | 8.5 | A2877$^{c}$ | 0.472 | | | | | |
| A1060$^{e}$ | 0.632 | 0.927 | 0.022 | 5 | 11 | 13.9 | A3158$^{e}$ | 0.667 | 0.804 | 0.035 | -72 | 4 | 8.5 |
| A1367$^{e}$ | | 0.584 | 0.050 | -35 | 4 | 12.8 | A3186$^{m}$ | 0.630 | 0.904 | 0.083 | -60 | 53 | 3.2 |
| A1644$^{e}$ | 0.448 | 0.893 | 0.031 | 69 | 11 | 6.4 | A3266$^{e}$ | 1.343 | 0.796 | 0.026 | 68 | 4 | 12.8 |
| A1650$^{e}$ | 0.788 | 0.807 | 0.029 | -6 | 5 | 5.3 | A3376$^{m}$ | 0.783 | 0.339 | 0.107 | 76 | 5 | 9.6 |
| A1656$^{e}$ | 0.760 | 0.806 | 0.016 | 80 | 2 | 14.9 | A3391$^{e}$ | 0.523 | 0.613 | 0.068 | 65 | 4 | 8.5 |
| A1689$^{e}$ | 0.833 | 0.888 | 0.023 | 26 | 7 | 5.3 | A3395SW | | 0.463 | 0.068 | -57 | 12 | 4.3 |
| A1736$^{e}$ | 0.580 | 0.852 | 0.040 | -82 | 22 | 9.6 | A3395NE | | 0.572 | 0.096 | -60 | 14 | 4.3 |
| A1767 | 0.614 | 0.859 | 0.050 | 24 | 17 | 6.4 | A3526$^{ec}$ | 0.395 | 0.832 | 0.018 | -89 | 2 | 9.6 |
| A1775 | 0.652 | 0.950 | 0.042 | -57 | 31 | 4.3 | A3532$^{e}$ | 0.573 | 0.791 | 0.069 | 42 | 35 | 5.3 |
| A1795$^{ec}$ | 0.674 | | | | | | A3667$^{e}$ | 0.537 | 0.601 | 0.033 | -53 | 2 | 7.5 |
| A1983$^{m}$ | 0.610 | 0.844 | 0.041 | -69 | 32 | 4.3 | AWM7$^{ec}$ | 0.460 | 0.665 | 0.089 | -83 | 7 | 7.5 |
| A2029$^{ec}$ | 0.685 | | | | | | CygA$^{ec}$ | 0.523 | 0.882 | 0.025 | -35 | 4 | 8.5 |
| A2052$^{ec}$ | 0.562 | 0.802 | 0.039 | 31 | 5 | 6.4 | MKW3S$^{ec}$ | 0.633 | 0.917 | 0.029 | 86 | 17 | 5.3 |
| A2063$^{ec}$ | 0.637 | | | | | | 3C129$^{e}$ | | 0.769 | 0.068 | 85 | 7 | 11.7 |

$^{c}$ central cooling time $< 10^{10}$years $\quad$ $^{e}$ member of Edge *et al.* sample

$^{\ddagger}$ $\beta$'s accurate to 17% $\quad$ $^{m}$ no measured X–ray temperature

The measurements on three clusters make the differences abundantly clear: (i) the X–ray emission in Abell 754 is very flattened near the core and becomes more circular outside the core ($\eta = 0.579$ versus $\eta_{MKU} = 0.820$); (ii) Abell 548 is more circular near the core than outside the core ($\eta = 0.886$ versus $\eta_{MKU} = 0.506$; and (iii) Abell 2052 has a roughly constant axial ratio ($\eta = 0.802$ versus $\eta_{MKU} = 0.807$).

## 2.5 ORIENTATION

We also extract the cluster orientation from the Fourier expansion of the photon distribution within the annulus.

$$\theta_{\bar{r}} = \frac{1}{2} \tan^{-1} \left( \frac{B_2}{A_2} \right) \tag{2.7}$$

We combine the formal and Monte–Carlo errors exactly as we do for the axial ratio.

Many clusters have orientations which vary as a function of radius. We combine the cluster orientation as a function of radius to produce an emission weighted average orientation $\theta_o$ within the $\langle \frac{S}{N} \rangle > 5$ region of the cluster image.

$$\theta_o = \left[ \sum_j (N\theta_{\bar{r}})_j \right]^{-1} \sum_j (N\theta_{\bar{r}})_j \tag{2.8}$$

where $N$ is the number of photons within each annulus. As with $w_{\vec{x}}$ and $\eta$, we use cluster simulations to test our accuracy; the third column of Figure 2 shows the test results for $\theta_o$. The top row in the column shows the absolute error of each measurement as a function of the measured cluster ellipticity. As expected, the orientation error is larger for more circular clusters; there is no evidence of a measurement bias ($\langle \Delta\theta \rangle = -0.85°$). The bottom row contains a distribution of scaled errors. The best fit Gaussian is centered near zero with an excess width of 49% ($\sigma_\theta$=1.49). Thus, we conclude that the orientation measurement is unbiased, but the uncertainties are underestimated. We increase all our orientation uncertainties by 49% to correct for this underestimate.

Columns labeled $\theta_o$ and $\sigma_\theta$ of Table 2 list the orientation angles and uncertainties for 59 clusters. Columns labeled $R[']$ contain the radius (in arcmin) within which we calculate the emission weighted average. Appendix A contains a detailed discussion of the individual measurements. Two clusters have variations in orientation or isophote twisting above the $3\sigma$ level. These two clusters are A426 ($\mathrm{RMS}_\theta = 33.1°$, $\sigma_{\mathrm{RMS}} = 2.4°$) and A119 ($\mathrm{RMS}_\theta = 37.0°$, $\sigma_{\mathrm{RMS}} = 9.7°$). In the case of an equilibrium system this twisting would be evidence of triaxiality; however, in these two cases there is evidence of departures from equilibrium. The centroid of the cooling flow region of A426 and the centroid of the outer region of the cluster are significantly different (see Table 1 and MFG93). In A119, the surface brightness profile is inconsistent with that produced by a single mass clump. Fabricant *et al.* (1993) show that a multi–clump model suggested by the galaxy distribution produces an X–ray surface brightness profile consistent with the IPC image.

Rhee & Latour (1991) analyze the X–ray orientations of 26 Abell clusters. They focus on the entire X–ray image, using the largest circular region centered on the peak of the X–ray emission which is not affected by the ribs. There are 11 measurements which appear in Rhee & Latour and Table 2. Although the measurement region generally differs slightly (Rhee & Latour typically include regions which do not satisfy our signal–to–noise constraint), the values generally agree; the scaled difference distribution $\nu = (\theta_0 - \theta_{RL})/\sigma$ has a mean $\langle \nu \rangle = -0.04$ and a width $\mathrm{RMS}_\nu = 1.2$. MKU measure orientations for 49 Abell clusters. Because of the different region used (see previous discussion), the orientations measured by MKU should be consistent with our measurements only in clusters where there is no isophote twisting.

## 2.6 RADIAL FALL–OFF: $\alpha$ AND $\beta$

There have been many attempts to understand the relationship between the radial distribution of the X–ray emitting gas and the radial distribution of galaxies (Abramopoulos

& Ku 1983; Jones & Forman 1984). Because both the gas and the galaxies respond to the same gravitational potential, the Euler equation for the gas and the Jeans equation for the galaxies imply a relationship between the radial distributions of the two cluster components in an equilibrium system. If the King approximation to the isothermal sphere describes the galaxy distribution ($\rho_{gal} \propto r^{-3}$ outside the core), $\beta$ describes the radial distribution of the gas ($\rho_{gas} \propto r^{-3\beta}$, where $\beta = \frac{kT}{\mu m_p \sigma^2}$, $T$ is the gas temperature, $\sigma$ is the galaxy velocity dispersion, and $\mu m_p$ is the mass of the average gas particle). Where the X–ray emissivity $\epsilon \propto \rho_{gas}^2$ (e.g. thermal bremsstrahlung), this $\beta$ prescription implies a cluster X–ray surface brightness profile of the form in Equation 2.3.

Some clusters in our ensemble are not well fit by this $\beta$ model. However, the radial behavior of the cluster is an important morphological parameter. Thus, we examine two consistent approaches to measuring the radial fall–off. We measure $\beta$ by first azimuthally averaging around the peak in the X-ray surface brightness to a radius of $16'$ (this includes the region within the IPC ribs for clusters located in the center of the field); we exclude point sources, the IPC ribs and outer regions, and separated subclumps (see Appendix A). We use the smoothed images to define this peak, but then produce the radial profile from the unsmoothed image (PSF has FWHM$\sim 1.5'$). The uncertainty in the surface brightness at a particular radius follows from the uncertainties in the image pixels which determine the value. We then find the parameter values which minimize the $\chi^2$ in the fit of the measured radial profile to a spherical $\beta$ model.

We require a confidence limit of 5% for the $\chi^2$ fit. In some clusters (often the ones with the highest quality images) this fit criterion is not met over the entire radial profile. This departure is sometimes caused by a cooling flow in the cluster core; previous $\beta$–fitting studies typically exclude these flow regions (Jones & Forman 1984). Because the X–ray image provides no clear boundary between the bright inner core and the outer regions, we define an objective core exclusion procedure. We exclude inner points in the radial profile *until* the fit to the remaining points has at least a 5% probability of consistency. Following this procedure we include fits to all clusters except those where (i) the core radius is comparable to the IPC field of view, (ii) the cluster is too complex to warrant radial averaging, or (iii) we exclude so many inner points to get an acceptable fit that the constraints on the fit parameters are unacceptably weak.

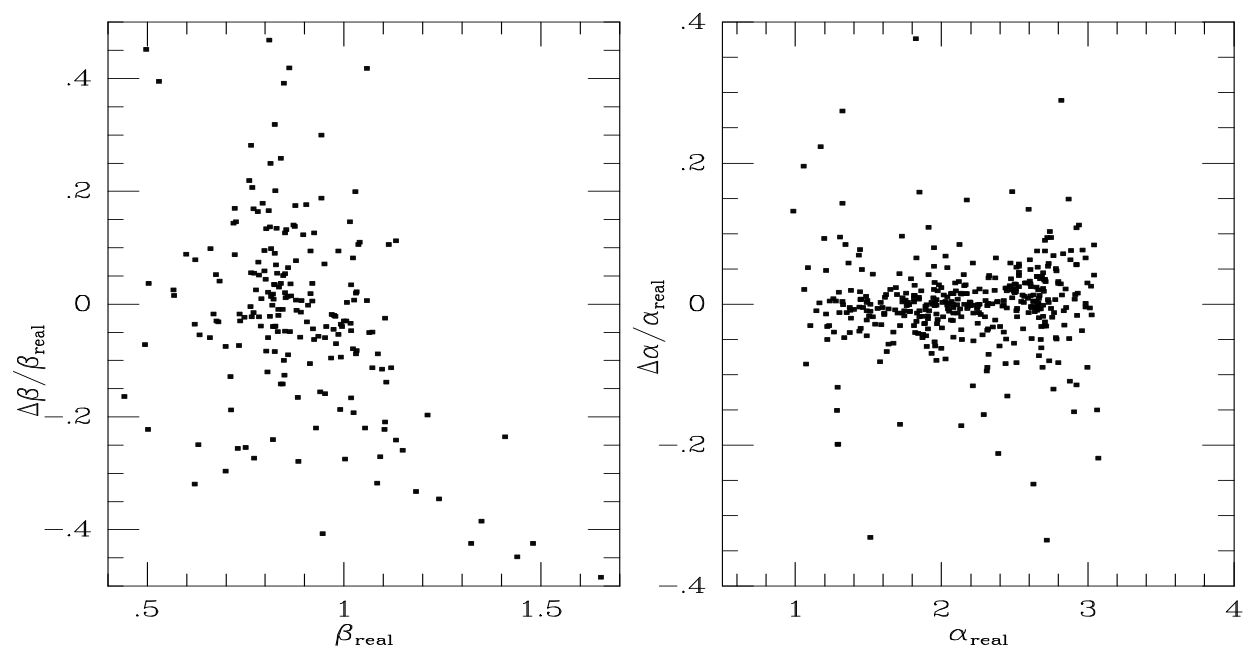

Figure 4: The accuracy of two different radial fall–off parameters, $\alpha$ and $\beta$: From left to right the plots are (i) the fractional error in $\beta$ measurements as a function of $\beta_{real}$ and (ii) the fractional error in the mean fall–off $\alpha$ versus $\alpha_{real}$. The width in the $\beta$ error distribution is $RMS_\beta = 17\%$ and the bias is $\langle \Delta\beta/\beta_{real} \rangle = -1.6\%$. The width in the $\alpha$ distribution is $RMS_\alpha = 7.5\%$ and the bias is $\langle \Delta\alpha/\alpha_{real} \rangle = 0.2\%$.

Here again we use the cluster simulations to test the accuracy of the $\beta$ measurements.

We compare the $\beta$ values for more than 200 pure images with the $\beta$ values for the artificial IPC images (Figure 4 column 1). Clearly, the uncertainty in the measured $\beta$'s is quite high; although there is no compelling evidence (except perhaps for $\beta > 1.2$) for a measurement bias ($\langle\Delta\beta/\beta_{real}\rangle = -1.6\%$), the scatter is quite large ($\mathrm{RMS}_\beta = 17\%$). This scatter is related to the correlation between the core radius $R_c$ and the radial fall–off $\beta$ (large $R_c$ favors steeper $\beta \rightarrow$ A3266 has a large $\beta$ probably because of the large core radius from an ongoing merger) and to the limited extent of the IPC field. The Poisson noise and instrumental imperfections affect $\beta$ by (i) slightly shifting the peak in the X–ray surface brightness which results in an increase in $R_c$ and (ii) introducing uncertainties in the radial profile which are largest for the lowest surface brightness (outermost) regions. We underestimate the $\beta$'s in the clusters with large $\beta_{real}$ in Figure 4 because the pure images are not well fit by the $\beta$–model. Therefore, the cores are excluded in achieving a decent fit to the idealized radial profiles. In the artificial IPC images of these clusters, the Poisson noise introduces large enough uncertainties in the radial profiles that the entire profile (including the core) is a good fit to the $\beta$–model in each case. So, in the pure images we have large $R_c$ and $\beta$, while in the artificial IPC images we have smaller $R_c$ and $\beta$. There is only one obvious fix for this problem: a genuinely accurate $\beta$ fit requires reasonable signal–to–noise in the surface brightness profile well outside the true core.

Even with the problems associated with $\beta$, we include measured $\beta$'s because of the historically attributed physical significance. (We do not include the core radii because there are additional problems associated with measuring them.) Columns labeled $\beta$ of Table 2 contain a list of 54 radial fall–off values (accurate to 17%). 48 of these values are for clusters with measured $L_X$ and $T_X$. The KS test comparison of this sample with the Edge *et al.* flux limited sample shows that the two samples are statistically indistinguishable ($T_X$: D=0.18 (40%), $L_X$: D=0.20 (25%)). Figure 3 contains a histogram of the $\beta$ values for this representative sample of clusters. The mean and width of this distribution are $\langle\beta\rangle = 0.65$ and $\mathrm{RMS}_\beta = 0.16$. The specifics of individual measurements are in Appendix A.

Jones & Forman (1984) determine the most likely range of $\beta$ for 46 clusters observed with *Einstein* and are currently completing work on a sample of more than 300 IPC cluster observations (Jones & Forman 1994). 24 $\beta$ measurements appear in both the Jones & Forman (1984) work and Table 2, and the agreement is generally good. Using the mean of the two Jones & Forman $\beta$'s as the best fit value and the difference as $\sim 4\sigma$, we calculate the offset between our values and their values $\langle\beta - \beta_{JF}\rangle = -0.03$. The width of the scaled difference distribution $\nu = (\beta - \beta_{JF})/\sigma$ is $\mathrm{RMS}_\nu = 0.75$. Some of the discrepancies between the two measurements is undoubtedly caused by differences in our cooling–flow exclusion strategy. Jones & Forman exclude inner bins until "a minimum in $\chi^2$ is reached", while we exclude bins until there is at least a 5% probability that the fit and data are consistent.

The large scatter in the $\beta$ measurements and the $\beta$–$R_c$ correlation lead us to search for an alternative method of comparing the mean fall–off within the same physical region of two different clusters. We fit the mean radial fall–off $\alpha$ within some specified cluster region.

Specifically, we fit a line in log space: $\log(I) = a - (2\alpha - 1)\log(R)$ where $a$ is the intercept and $\alpha$ is the slope. Note that although a power law is not generally a good fit to the cluster radial profile (near the core), fitting a power law over a particular region (which can include the core) is a reasonable method of determining the *mean* fall–off in the surface brightness profile. Figure 4 contains a test of this approach using more than 200 artificial cluster images. The fractional error is plotted against the true value. We find no bias ($\langle \Delta\alpha/\alpha_{real} \rangle = 0.2\%$) and the scatter in $\alpha$ ($\mathrm{RMS}_\alpha = 7.5\%$) is much smaller than the scatter in $\beta$. Because of the improved scatter, we use $\alpha$ in §3 to directly compare the radial behavior of cluster ensembles over the same physical scale.

## 3. THE MORPHOLOGY–COSMOLOGY CONNECTION

We use our ensembles of observed and simulated clusters to constrain three preferred cosmological models. For each cosmological model we ask, "Is it possible for the clusters we observe today to develop in a universe described by this cosmological model?" Because it is impossible to remove the observational effects from the observed clusters, we introduce these effects into the simulated data. We construct an ensemble of 24 cluster simulations (8 sets of initial conditions evolved in three different models– the same simulations discussed in Evrard *et al.* 1993) from which we extract artificially observed ensembles similar to the observed cluster ensemble.

Our observed sample consists of 52 clusters with measured X–ray temperatures and luminosities. We exclude 5 clusters because of their low redshifts (A426, A3526, A1060, AWM7, and A1367). We first describe the simulations in detail and then turn to the procedure for comparing the simulated clusters with the observed clusters. A discussion of the results and the cosmological constraints follows.

### 3.1 THE CLUSTER SIMULATIONS

We employ a set of 24 simulations comprised of three sets of eight random density fields. The three sets correspond to three different assumptions for the underlying cosmology. The models we investigate are: (i) a biased, Einstein–deSitter universe with normalization $\sigma_8 = 0.59$ (where $\sigma_8 \equiv \langle (\delta\rho/\rho)^2 \rangle^{1/2}$ on an $16 h_{50}^{-1}$ Mpc scale); (ii) an unbiased ($\sigma_8 = 1.0$), open universe with $\Omega_o = 0.2$ and (iii) an unbiased, low density model with $\Omega_o = 0.2$ and $\lambda_o = 0.8$.

A cold dark matter power spectrum with $\Gamma = \Omega h = 0.5$ is used to generate initial displacement fields on a $32^3$ spatial grid as in Efstathiou *et al.* (1985). The periodic grid models a comoving cube of space with length $L$. To sample a range of cluster richness, we evolve two random realizations from four different box lengths $L = 30$, 40, 50 and 60 Mpc. The initial density field is constrained using the method of Bertschinger (1987) in order to form a significant cluster within the volume. The applied constraint is that the density field $\delta$ at the center of the volume, when filtered with a Gaussian spatial filter $W(r) \propto \exp -r^2/2r_f^2$ with $r_f = 0.2L$, take on a present, unbiased, linearly evolved value of $\delta_o = 3.0$. This corresponds to perturbations $2.5 - 5$ times the RMS amplitude over the range of box sizes.

The P3MSPH algorithm (Evrard 1988) is used to propagate the initial, linear density fields to their present, deeply non–linear states. Two sets of $32^3$ particles, coupled by gravity, represent the dark matter and baryonic fluids. The baryons are approximated as an ideal, $\gamma = 5/3$ gas with constant mean molecular weight $\mu = 0.6$ appropriate for a fully ionized, primordial plasma. The thermal energy changes adiabatically and as a result of shocks generated in regions of converging flow. A cosmic baryon fraction $\Omega_b = 0.1$ is assumed for all the cosmologies. This approach allows for some amount of collisionless dark matter in each model.

Gravity is softened with a pairwise, Plummer potential $\Phi(r) = -1/(r^2 + \varepsilon^2)^{1/2}$ (in units where $G = m = 1$) with $\varepsilon = 0.0023L$ kept fixed in the comoving frame. The smoothed particle hydrodynamic (SPH) method entails the use of a smoothing kernel $W(r, h_i)$ which we take to be a Gaussian $W(r, h_i) = (\pi h_i)^{-3/2} \exp -r^2/h_i^2$. The hydrodynamic smoothing length $h_i$ for particle $i$ is varied to keep

$$\frac{4\pi}{3}\rho_i h_i^3 \;=\; c_1 \tag{3.1}$$

with $c_1 = 8\pi$. Because 95% of the kernel's weight falls within $2h_i$, there are roughly 200 particles within the domain of hydrodynamic influence of each particle. It is straightforward to show that the Gaussian kernel is similar to the B-spline employed by Hernquist & Katz (1989) and others with a modest rescaling of $h$ (Evrard, Feldman & Watkins 1994). The smoothing length cannot increase beyond a value $h_{max} = 0.025L$, a value derived as a fixed fraction of the P3M short range force cell size. An absolute lower bound on $h_i$ of $0.5\varepsilon$ is imposed. In addition, each particle's value of $h_i$ is subject to a lower bound derived from the Courant condition (see Evrard 1988). In practice, the timestep is set small enough and the mass resolution is modest enough that this bound does not come into play. The minimum $h_i$ achieved is comparable to $\varepsilon$. The 'effective' resolution of the runs is approximately $0.005L$.

The initial displacement fields are normalized so that a growth factor of 16 is required to recover the original, unbiased spectrum. For the biased, $\Omega$=1 models, the initial state corresponds to a redshift $z_i = 9$. For the low density models, the initial state corresponds to redshifts $z_i = 47.5$ ($\Omega_o$=0.2) and 23.0 ($\Omega_o$=0.2 & $\lambda_o$=0.8). The code uses a fixed timestep to advance the particles, with 750, 1400 and 900 steps used for the $\Omega$=1, $\Omega_o$=0.2 and $\Omega_o$=0.2 & $\lambda_o$=0.8 runs, respectively. Energy in the Layzer–Irvine equation (Efstathiou *et al.* 1985) is conserved to typically better than 1% for the $\Omega$=1 runs and $\sim$ 5% for the low density models. Because the difference equations for the gas are not accurate to second order, errors in the baryonic energy are larger than those of the dark matter.

### 3.2 COMPARING OBSERVED AND SIMULATED CLUSTERS

For each *Einstein* cluster observation we extract a similar observation of a simulated cluster from each of the cosmological models (see Figure 5). Specifically, for each of the 52 *Einstein* clusters we (i) choose one of the eight sets of initial conditions, (ii) scale the three (one for each cosmological model) present epoch configurations for the particular initial conditions so that the X–ray emitting gas in the three simulations has the same temperature as the gas in the observed cluster, (iii) choose a random observing direction and then (iv) image each present epoch configuration (Evrard 1990) at the same redshift as the observed cluster (for $H_0$ = 50 km s$^{-1}$Mpc$^{-1}$). The scaling conserves the mean density within the simulation volume while changing the temperature: $T \rightarrow \gamma T$ subject to constant $M/R^3$ implies the scale change $R \rightarrow \sqrt{\gamma} R$. For the entire ensemble of simulated images, the spatial scaling $\sqrt{\gamma}$ varies from 0.6 to 2.9 with median values 1.1 for the $\Omega$=1 models, 1.2 for the $\Omega_o$=0.2 models, and 1.5 for the $\Omega_o$=0.2 & $\lambda_o$=0.8 models. Because of the differences in cluster X–ray luminosity (for the default scaling, the low $\Omega_o$ models produce more luminous clusters), we select the observation time for each model so that the *number of photons* in the region of interest is identical for all four ensembles. In this way we produce four ensembles (1 observed and 3 artificial) of cluster observations which are homogeneous in three respects; each ensemble has identical distributions in (i) gas temperature, (ii) spatial scale, and (iii) number of cluster photons within the region of interest. This approach isolates the differences in cluster morphology.

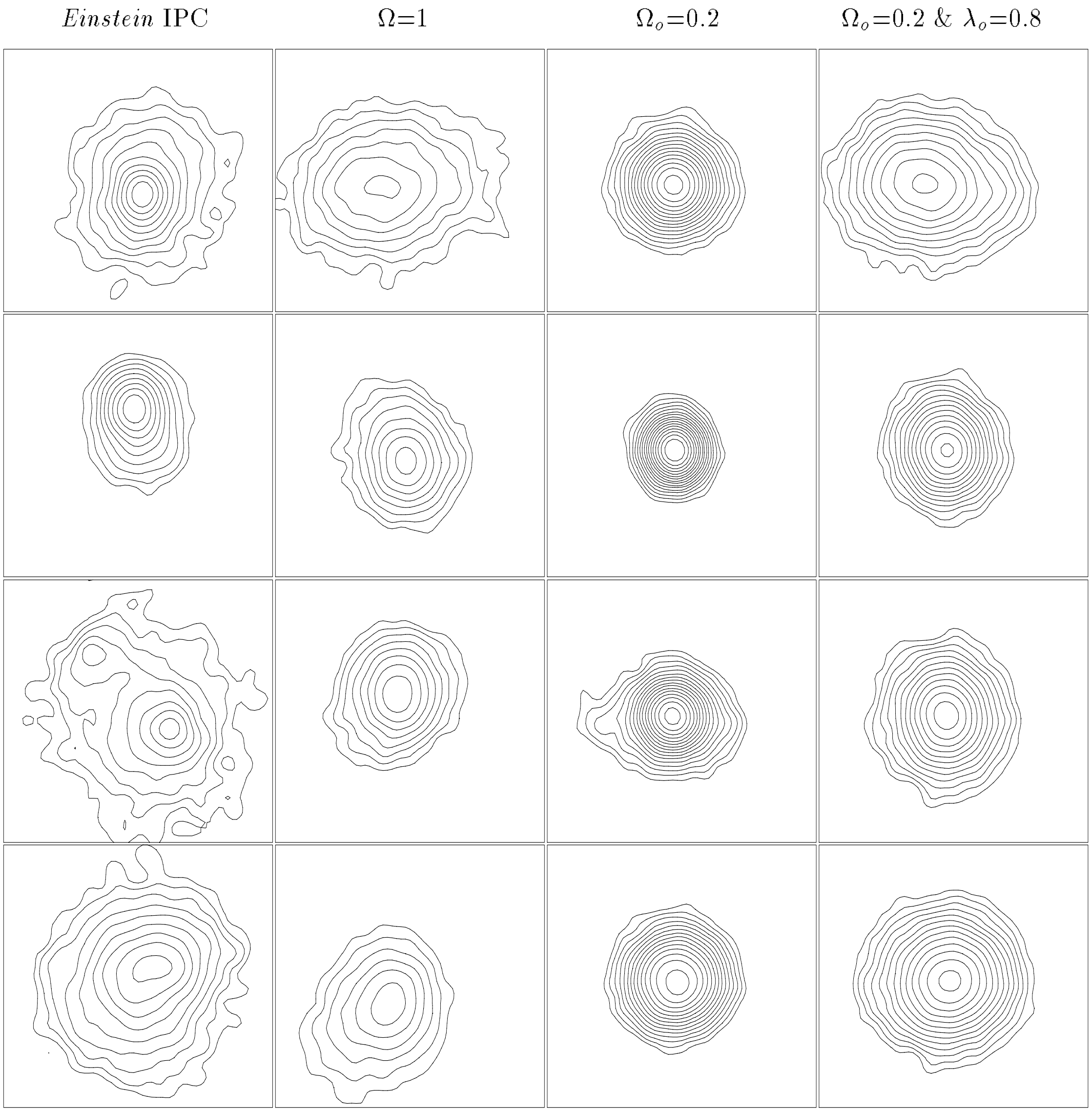

Figure 5: Contour maps of *Einstein* IPC images of 4 Abell clusters and the artificial IPC images of cluster simulations. From the left to right the columns are (i) *Einstein* IPC images, (ii) artificial IPC images of $\Omega$=1 cluster simulations, (iii) artificial images of $\Omega_o$=0.2 simulations, and (iv) artificial images of $\Omega_o$=0.2 & $\lambda_o$=0.8 simulations. From top to bottom the clusters are (i) Abell 496, (ii) Abell 644, (iii) Abell 1644, and (iv) Abell 2256. In producing the artificial IPC images of cluster simulations, we image the simulations at the same redshift as the associated Abell cluster. Contours are spaced by factors of 1.5 in surface brightness. For each image, the surface brightness of the bottom contour is the mean surface brightness in the outermost annulus which satisfies the $\langle \frac{S}{N} \rangle > 5$ constraint.

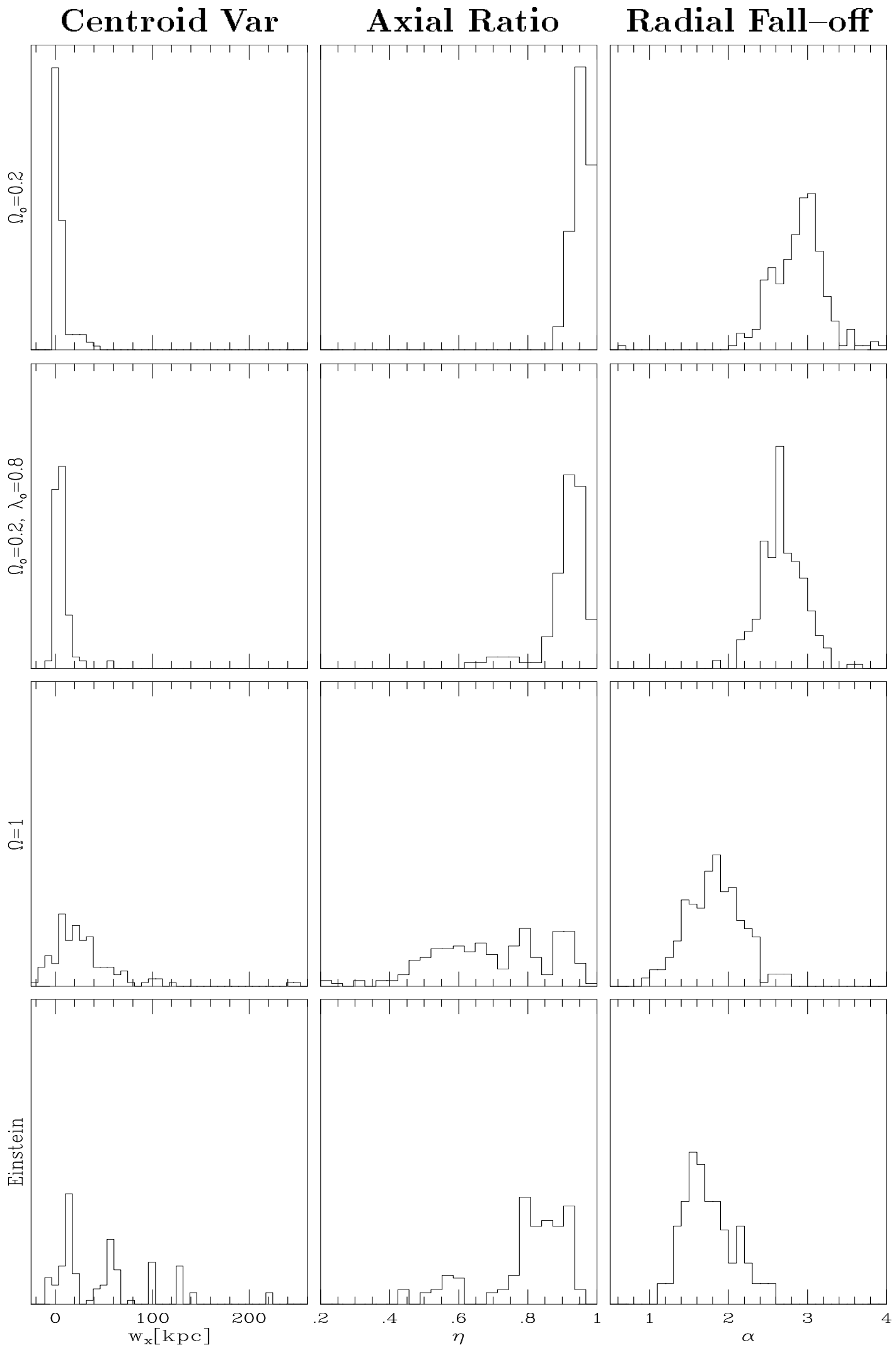

Figure 6: Comparing X–ray Morphologies. From left to right each column contains normalized distributions of (i) the emission weighted centroid variation $w_{\vec{x}}$ in $h_{50}^{-1}$kpc (ii) the emission weighted axial ratio $\eta$, and (iii) the radial fall–off $\alpha$. From top to bottom the rows correspond to the following cosmological models: (i) $\Omega_o$=0.2, (iii) $\Omega_o$=0.2 & $\lambda_o$=0.8, and (iii) $\Omega$=1. The bottom row is the *Einstein* IPC sample. Table 3 lists the mean and width of these distributions.

To increase the sample size and fully probe the morphological characteristics of the simulations, we cycle through the 52 *Einstein* IPC observations 5 times, producing a total of 260 different comparison–sets of cluster observations. Each set consists of one *Einstein* IPC observation and artificial observations of the same initial density field evolved in each of the three cosmological models (note that because we only use 52 *Einstein* images, each of the IPC images appears 5 times). We then reduce and analyze the artificial images as though they were normal IPC observations. We compare the X–ray morphologies of the 4 ensembles of 260 images using three different parameters: the emission weighted centroid variation $w_{\vec{x}}$, the emission weighted axial ratio $\eta$, and the radial fall–off $\alpha$. Figure 6 contains the distributions in these three parameters. In making these comparisons we treat each set of 4 images as a unit to preserve the ensemble similarities in $T_X$, $z$, and photon number; if making a measurement on any one of the 4 images is impeded then we exclude the entire set (For example, from Table 1 it is clear that A2142 has no $w_{\vec{x}}$ measurement because of a central point source. Rather than just excluding the cluster $w_{\vec{x}}$ for A2142, we exclude all 5 A2142 comparison sets–4 images each.).

## 3.3 THE DISTRIBUTION OF CENTROID VARIATION

The first column of Figure 6 contains the distribution in $w_{\vec{x}}$ for the four samples of cluster images. For the observed clusters and the $\Omega$=1 simulations we calculate $w_{\vec{x}}$ starting from an inner radius of $4'$, but for the $\Omega_o$=0.2 and $\Omega_o$=0.2 & $\lambda_o$=0.8 simulations we exclude those innermost annuli with radii smaller than the scale of one "effective" resolution length. This is necessary because the cluster cores in the low $\Omega_0$ models have evolved to central densities so high that corresponding cooling times are less than a Hubble time. Because the present simulations do not include radiative cooling, the innermost cores of the low $\Omega_0$ simulations are non–physical and must be excluded. (Calculating $w_{\vec{x}}$ with these bright cores included tends to increase the differences between the $w_{\vec{x}}$'s measured in the $\Omega$=1 model and the low $\Omega_0$ models.) Clearly, clusters with large centroid variations are much more probable in the observed sample and the $\Omega$=1 model than in the two low $\Omega_0$ models. KS test results indicate that the $\Omega$=1 model and the observed sample are marginally consistent, and that the two low $\Omega_0$ distributions are marginally consistent. All other combinations differ significantly.

Because a sizable centroid variation is a signature of dynamical youth (MFG93, Mohr & Evrard 1994), it is no surprise that these distributions reflect the differences in the merger histories of the three models. As predicted by Richstone, Loeb, & Turner (1992), the clusters evolving in the $\Omega$=1 model are more likely to experience significant growth at late times than those evolving in the low $\Omega_0$ models. The similarity between the $\Omega$=1 distribution and the observed cluster distribution suggests similarities in the evolutionary histories of these two ensembles: both the observed and the simulated clusters have a high probability of late–time growth.

## 3.4 THE DISTRIBUTION OF AXIAL RATIO

The second column of Figure 6 shows the striking differences in the distributions of $\eta$. As with the centroid variations, we exclude one "effective" resolution length from the cores of the low $\Omega_0$ clusters. The absence of late–time mergers in the two low $\Omega_o$ models produces predominantly spherical clusters at the present epoch ($\Omega_o$=0.2: $\langle\eta\rangle$ = 0.95; $\Omega_o$=0.2 & $\lambda_o$=0.8: $\langle\eta\rangle$ = 0.91). The higher probability of late–time mergers in the $\Omega$=1 models results in more highly flattened clusters thus reducing the average axial ratio ($\langle\eta\rangle$ = 0.70). The sample of *Einstein* clusters ($\langle\eta\rangle = 0.80$) lies intermediate between the $\Omega$=1 model and the two low $\Omega_o$ models. A KS test indicates that all four distributions differ significantly.

The physics not included in the numerical simulations (e.g. gas cooling) will undoubtedly affect the distributions of $\eta$. For example, no cooling flows can form in the simulations, but observed cooling flows are characterized by circularly symmetric peaks in the X–ray surface brightness. Such a flow tends to weight the distribution of axial ratio toward 1 (especially for an emission weighted axial ratio). We examine this effect using X–ray images of observed clusters (see Figure 3). Of the 51 clusters included in the Figure 3 $\eta$ distribution, 44 have been analyzed for the presence of cooling flows (Edge, Stewart, & Fabian 1992, Stewart *et al.* 1984). 13 of these 44 clusters have central cooling times which are significantly less than $10^{10}$ years (these clusters are marked with a $^c$ in Table 1 and Table 2). The dashed lines in Figure 3 show the $\eta$ distribution after the removal of these cooling flow clusters. With one exception, the clusters containing definite cooling flows fall in the region $\eta > 0.8$ (the region with fewer clusters in the $\Omega$=1 model).

On the other hand, recent cluster mergers tend to weight the distribution toward the highly flattened end. The simulations do follow cluster mergers properly, so including cooling in the simulations would shift the $\eta$ distribution toward 1. Such a shift would make the $\Omega$=1 model more consistent with the observed clusters. Because the $\Omega_o$=0.2 and $\Omega_o$=0.2 & $\lambda_o$=0.8 models have $\eta$ distributions which are already more spherical than observed clusters, adding cooling would not significantly improve their match to observed clusters.

Table 3. Mean (and RMS) of $w_{\vec{x}}$, $\eta$, and $\alpha$ Distributions

| Parameter | *Einstein* | $\Omega$=1 | $\Omega_o$=0.2 & $\lambda_o$=0.8 | $\Omega_o$=0.2 |
|---|---|---|---|---|
| $w_{\vec{x}}$[kpc] | 50.1 (49.2) | 30.4 (39.3) | 6.6 (8.8) | 5.4 (7.9) |
| $\eta$ | 0.80 (0.12) | 0.70 (0.17) | 0.91 (0.07) | 0.95 (0.02) |
| $\alpha$ | 1.75 (0.32) | 1.82 (0.36) | 2.68 (0.27) | 2.88 (0.36) |

## 3.5 THE DISTRIBUTION OF RADIAL FALL-OFF

The third column in Figure 6 contains the normalized $\alpha$ distributions for the four ensembles. The fit region for $\alpha$ is determined separately for each of the 260 comparisons. In each case the inner boundary is the larger of the "effective" resolution lengths for the $\Omega_o$=0.2 and $\Omega_o$=0.2 & $\lambda_o$=0.8 models; the same core region is excluded from all four clusters within each of the 260 comparisons. As with the $\eta$'s and $w_{\vec{x}}$'s, the radial fall–off $\alpha$ neatly separates the low $\Omega_0$ simulations from the $\Omega$=1 models and the observed clusters. The radial profiles of the low $\Omega_0$ clusters are significantly steeper. A KS test indicates that the $\alpha$ distributions of the $\Omega$=1 clusters and observed clusters are marginally consistent.

It is not possible to make the low $\Omega_0$ clusters consistent with observed clusters by changing the slope of the power spectrum (Crone, Evrard, & Richstone 1994); however, gas physics excluded from the simulations may significantly affect the cluster radial fall–off. Metzler & Evrard (1994) investigate the effects of gas ejection from galaxies in a series of simulations. They produce an ensemble of initial density perturbations consistent with standard, $\Omega$=1

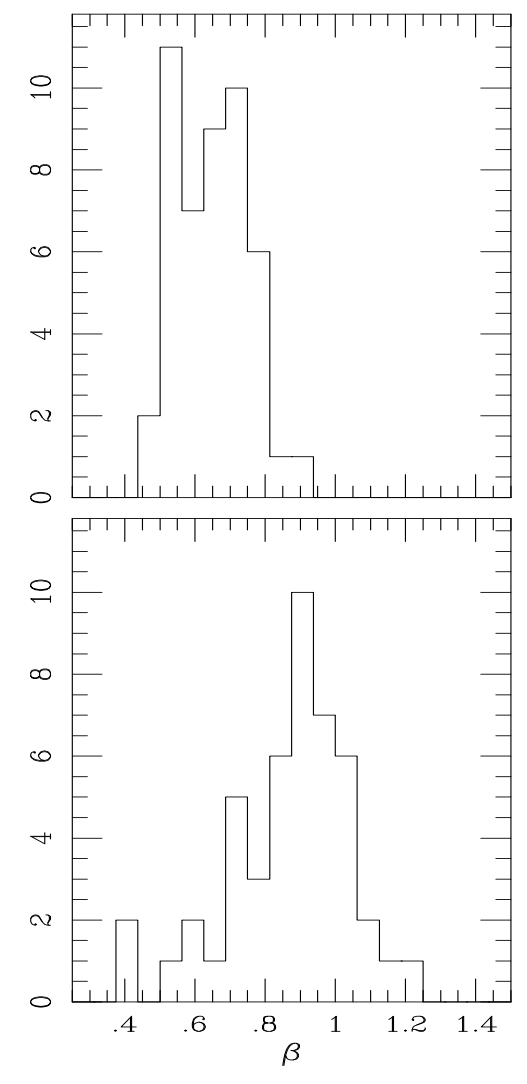

Figure 7: We demonstrate the effects of ejection from galaxies on the cluster radial fall–off using cluster simulations which include gas ejection (Metzler & Evrard, 1994). The histograms contain the results of $\beta$–fits to idealized images of 47 different clusters. The top histogram is the radial fall–off for those clusters evolved with ejection from galaxies ($\langle\beta\rangle = 0.65$), and the bottom histogram contains the radial fall–off for the identical clusters evolved without ejection ($\langle\beta\rangle = 0.87$).

CDM constrained to form clusters in a range of box sizes. They then evolve each set of these initial conditions twice. In the first set of simulations they simply evolve the baryonic and dark matter components as described in §3.1. In the second set of simulations they investigate ejection from galaxies by (i) replacing $2.5\sigma$ density peaks on galaxy scales ($5\times10^{11}\mathrm{M}_\odot$) with single, massive, dark matter particles and (ii) modeling continuous gas ejection from galaxies with the discreet ejection of hot gas particles from galaxies over time. They explore the maximal effect of ejection from galaxies by following an extreme ejection model: constant ejection totalling half the galaxy mass from $z$=4 to the present epoch. In Figure 7 we plot the histograms of radial fall–off $\beta$ for the clusters evolved without ejection (bottom) and those evolved with ejection (top). The clusters evolved with no ejection ($\langle\beta\rangle = 0.87$) have gas distributions which fall off more steeply than the same clusters evolved with ejection ($\langle\beta\rangle = 0.65$); the $\beta$ distribution of the clusters evolved with ejection is very similar to the intrinsic $\beta$ distribution of observed clusters (see Figure 3).

The Metzler & Evrard simulations demonstrate that gas ejection can significantly decrease the radial fall-off of the ICM density. Since the models employ a fairly extreme ejection history for galaxies, a more realistic case would introduce a smaller effect. Figure 6 shows the difference in the average value of the mean slope parameter $\alpha$ between the observations and models is $\sim$5% for $\Omega = 1$ and $\sim$40% for low $\Omega_0$. Ejection at a level higher than that employed by Metzler & Evrard may be advocated to make the profile slopes for the low density runs consistent, while a more modest level of energy input from galaxies would bring the $\Omega = 1$ runs into agreement with the observations. At any rate, because the bulk of ejection in the low $\Omega_o$ models would occur at very high redshift, it would not be able to alter the isophote shifts or axial ratios of the present day X–ray images. The low $\Omega_0$ models would still disagree with the observations on these measures.

## 3.6 THE EMISSION WEIGHTING AND CORE EXCLUSION

To ensure that the morphological differences between the low $\Omega_0$ clusters and the $\Omega$=1

clusters are properly reflected in Figure 6, we test whether our methods of emission weighting and core exclusion (in the low $\Omega_0$ clusters) introduce measurement bias. To test emission weighting, we calculate $w_{\vec{x}}$ and $\eta$ for the entire ensemble of $\Omega_o$=0.2 & $\lambda_o$=0.8 and $\Omega$=1 simulations both with and without emission weighting. Specifically, we set $N = 1$ in Equation 2.2 and Equation 2.6 and recalculate the values. The comparison of more than 400 measurements for the $\Omega$=1 and $\Omega_o$=0.2 & $\lambda_o$=0.8 simulations reveals that the mean change in axial ratio $\eta$ is 0.2% for the $\Omega_o$=0.2 & $\lambda_o$=0.8 clusters and 0.5% for the $\Omega$=1 clusters. The mean change in $w_{\vec{x}}$ is 0.016 arcmin for the $\Omega_o$=0.2 & $\lambda_o$=0.8 clusters and 0.020 arcmin for the $\Omega$=1 clusters (compared to the median uncertainty for the *Einstein* cluster sample of 0.083 arcmin). Clearly these changes are insignificant when compared to the large differences reflected in the Figure 6 histograms.

Measurements on the low $\Omega_0$ clusters are bounded on the interior by the unresolved core and on the exterior by the $\left\langle \frac{S}{N} \right\rangle > 5$ constraint. To determine whether $w_{\vec{x}}$ and $\eta$ are sensitive to the exact size of the excluded core, we double the core and compare the measurements. The artificial IPC image produced from each simulated cluster is exposed to obtain the same number of cluster photons; therefore, doubling the diameter of the exclusion region increases the $\Omega_o$=0.2 & $\lambda_o$=0.8 observing times. Because of the increased observing times, the $\left\langle \frac{S}{N} \right\rangle > 5$ constraint is satisfied to larger radii. We find that for the $\Omega_o$=0.2 & $\lambda_o$=0.8 simulations, increasing the size of the core exclusion region by a factor of two results in a mean change in $w_{\vec{x}}$ of 0.026 arcmin and a mean change in $\eta$ of 0.5%. Clearly, the morphological measurements for this ensemble of low $\Omega_0$ clusters are not significantly affected by our core exclusion procedure.

## 4. CONCLUSION

We assemble a representative sample of 65 *Einstein* IPC cluster images to quantify the range of cluster X–ray morphologies. For each cluster we measure the emission weighted centroid variation ($w_{\vec{x}}$), the emission weighted axial ratio ($\eta$), the emission weighted orientation ($\theta_o$), and the radial fall–off parameters, $\alpha$ and $\beta$. We employ a Monte–Carlo approach to account for the effects of Poisson noise, instrumental imperfections, and foreground/background X–ray point sources. With the cluster simulations described in §3 we evaluate the measurement accuracy (Figure 2 & Figure 4). We list the measurements on each cluster (Table 1 & Table 2) and construct histograms (Figure 3) of the intrinsic distributions in $w_{\vec{x}}$, $\eta$, and $\beta$. This set of morphological measures is sensitive to the cosmological density parameter $\Omega$. The $w_{\vec{x}}$ measurements indicate that from 50% to 70% of clusters exhibit core substructure, depending on the adopted confidence limit.

We use the X–ray morphologies of observed and simulated clusters to place cosmological constraints. With the combined N–body/SPH code P3MSPH (Evrard 1988), we evolve eight different initial density fields (sampled from an $\Omega h = 0.5$ CDM spectrum) according to 3 different underlying cosmological models: (i) $\Omega$=1, $\sigma_8$=0.59; (ii) $\Omega_o$=0.2, $\sigma_8$=1; and (iii) $\Omega_o$=0.2 & $\lambda_o$=0.8, $\sigma_8$=1. These 24 simulations are the same ones analyzed by Evrard *et al.* (1994). Following a comparison method designed to isolate the cosmological parameters, we use 52 of the *Einstein* cluster images to compare the range of observed and simulated cluster morphologies. We demonstrate that (i) $w_{\vec{x}}$, $\eta$, and $\alpha$ neatly separate clusters evolved in low $\Omega_0$ models from those evolved in $\Omega$=1 models (see also Evrard *et al.* 1994), and (ii) clusters like those we observe today are very rare in low $\Omega_0$ universes.

How general are these results? Can any low $\Omega_0$ model reproduce the observed complexity of the X–ray images of real clusters? For the case of bottom–up Gaussian initial density fields, we suspect the answer is no. The reason is that merging is the most natural mechanism capable of generating asymmetric isophote structure in the ICM, and the present merging frequency of clusters is much more sensitive to the density parameter than to the shape of the power spectrum on cluster scales (e.g. Lacey & Cole 1993). Increasing $\Omega_0$ is the most natural way to increase the merger frequency and thereby produce a higher fraction of distorted X–ray images. This point is sharpened by the fact that we do not have complete freedom in specifying the shape of the power spectrum on cluster scales. The abundance of clusters as a function of X–ray temperature implies an effective spectral index $n$ between $-1$ and $-2$ (Henry & Arnaud 1991), and this range is consistent with power spectra derived from large–scale galaxy catalogues (Vogeley *et al.* 1992; Fisher *et al.* 1993; Feldman, Kaiser & Peacock 1994).

How might the low $\Omega_0$ models be saved? We have seen that energy ejection from galaxies can flatten the gas density profile. However, it is unclear whether a realistic level of ejection can produce an effect large enough to bring the $\alpha$ and $\beta$ measurements into agreement with observations. One might attempt to fix the axial ratios for the low density models by appealing to a higher dark matter fraction (presently 50% in our models) which could make the underlying cluster potentials more aspherical. However, the clusters would still be dynamically old, so the isophotes, even if less spherical, would retain their symmetric nature. Thus the centroid variation $w_{\vec{x}}$ would continue to present a strong constraint. To address these questions directly, we intend to pursue experiments of specific low density models in the near future.

Other recent attempts to constrain $\Omega_0$ from cluster properties also disfavor low density models. Matching the mean velocity dispersion of rich clusters with $\Omega_0 = 0.2$ requires $\sigma_8 \simeq 1.25 - 1.58$, implying galaxies are anti–biased with respect to the total mass content of the universe (Lilje 1992; White, Efstathiou & Frenk 1993). The strong increase in the fraction of blue galaxies in moderate redshift clusters (the "Butcher–Oemler effect") is also difficult to explain in a low density universe (Kauffmann 1994). A recent paper by de Theije, Katgert & van Kampen (1994) presents a possible exception to this trend. They find the ellipticities of the galaxy distributions within Abell clusters to be somewhat rounder than those found in their $\Omega = 1$ simulations. However, they did not yet have a complete set of low $\Omega_0$ simulations to compare against the observations. In general, any evidence for recent evolution or merging in the cluster population tends to disfavor low $\Omega_o$. With improved optical and X–ray data combined with detailed numerical simulations, specific cases of recently merged clusters are becoming more apparent. Examples include A2256 (Fabricant, Kent, & Kurtz 1989; Briel &

Henry 1994), A754 (Fabricant *et al.* 1986, Zabludoff & Zaritsky 1994) and Coma (Burns *et al.* 1994).

## ACKNOWLEDGEMENTS

We are very grateful to Chris Metzler for allowing us to use his ejection simulations. Thanks to Rob Hewett & Lisa Paton of the Computation Facility at the CfA who allowed my data to monopolize the public pool disks and often turned a blind eye to the countless simulations running in the background. Thanks also to the support staff of the Einstein Data Archive at the CfA for their patient responses to my many questions. This work is supported in part by the NSF graduate fellowship program, the NASA graduate fellowship program (NGT-51336), NAGW-201, NAGW-2367, and the Smithsonian Institution.

# APPENDIX A

In this appendix we list the details associated with measurements on specific *Einstein* IPC cluster images. Reductions for those clusters not listed follow the standard procedure described in the text. We define a set of guidelines and attempt to apply them uniformly to all cluster images. In measuring the cluster radial fall–off, these guidelines include (i) excluding well separated subclumps from the radial profile where possible, (ii) avoiding radial averaging for those clusters which are very complex, (iii) excluding point sources from the radial profile, and (iv) truncating the radial profile well before the IPC ribs. For $w_{\bar{x}}$ these guidelines are (i) excluding regions with bright point sources (the fainter, less obvious point sources are accounted for with a Monte–Carlo procedure), (ii) including subclumps except in those cases where there is independent evidence that the subclump is a chance superposition and not dynamically related to the main cluster, and (iii) avoiding those clusters where the $\langle \frac{S}{N} \rangle > 5$ region is so small that fewer than 4 centroids are measured. For $\eta$ and $\theta_o$ we (i) avoid regions containing bright point sources and (ii) exclude well separated subclumps. Angular distances and directions are given with respect to the peak in the X–ray surface brightness.

A85: $\alpha/\beta$: subclump to S excluded. $\eta/\theta_o$: Truncated at $R = 5.4'$ to exclude subclump. $w_{\bar{x}}$: subclump included. An extensive radial velocity study (Malumuth *et al.* 1992) provides evidence that 7 galaxies are associated with a foreground group; however, the spatial distribution of these 7 galaxies is not well centered on the subclump in the X–ray emission. Therefore, we consider the subclump to be evidence of an ongoing merger and therefore include it in the $w_{\bar{x}}$ measurement.

A133: $\alpha/\beta$: three point sources excluded (18′ N, 13′ SW, and 18′ S). $\eta/\theta_o/w_{\bar{x}}$: $\langle \frac{S}{N} \rangle > 5$ region does not include these three point sources.

A168: $\alpha/\beta$: cluster too complex for radial analysis. $\eta/\theta_o/w_{\bar{x}}$: no rib or point source contamination within $\langle \frac{S}{N} \rangle > 5$ region.

A400: $\alpha/\beta$: point source 15′ S of cluster excluded. No $\beta$ value included because core radius from $\beta$–fit larger than IPC field of view. $\eta/\theta_o/w_{\bar{x}}$: point source 15′ S of cluster is outside $\langle \frac{S}{N} \rangle > 5$ region.

A426: $\eta/\theta_o/w_{\bar{x}}$: truncated at $R = 13.9'$ to avoid rib contamination.

A478: $\alpha/\beta$: excluded two point sources (10′ NW and 25′ NE). $\eta/\theta_o/w_{\bar{x}}$: no point sources within $\langle \frac{S}{N} \rangle > 5$ region.

A539: $\alpha/\beta$: excluded one point source (3′ W/NW). $\eta/\theta_o/w_{\bar{x}}$: point source inhibits measurements.

A592: $\alpha/\beta$: truncated at $R = 13.3'$. $w_{\bar{x}}$: $\langle \frac{S}{N} \rangle > 5$ region too small for measurement.

A754: $\alpha/\beta$: point source excluded (25′ SW). No $\beta$ included because core radius is comparable to IPC field of view. $\eta/\theta_o/w_{\bar{x}}$: no point source within $\langle \frac{S}{N} \rangle > 5$ region.

A780: $\alpha/\beta$: point source excluded (17′ S/SE). $\eta/\theta_o$: point source outside $\langle \frac{S}{N} \rangle > 5$ region. $w_{\bar{x}}$: $\langle \frac{S}{N} \rangle > 5$ region too small to make measurement.

A1060: $\eta/\theta_o/w_{\bar{x}}$: truncated at $R = 13.9'$ to avoid rib contamination.

A1367: $\alpha/\beta$: IPC field of view inadequate to measure $\beta$. $\eta/\theta_o/w_{\bar{x}}$: point source near peak in surface brightness removed. Truncated at $R = 12.8'$ to avoid rib contamination.

A1644: $\alpha/\beta$: excluded subclump to NE. $\eta/\theta_o$: truncated at $R = 6.4'$ to exclude subclump. $w_{\bar{x}}$: subclump included. An extensive radial velocity study failed to detect a superposed subclump; therefore we consider the subclump to be evidence of an ongoing merger.

A1650: $\alpha/\beta$: one point source excluded (20′ SE). $w_{\bar{x}}$: $\langle \frac{S}{N} \rangle > 5$ region too small for measurement.

A1656: $\eta/\theta_o/w_{\bar{x}}$: Analysis truncated at $R = 14.9'$ to avoid rib contamination.

A1689: $w_{\bar{x}}$: $\langle \frac{S}{N} \rangle > 5$ region too small for measurement.

A1736: $\alpha/\beta$: excluded two point sources (17′ SE and 30′ E). $\eta/\theta_o/w_{\bar{x}}$: point sources are exterior to the $\langle \frac{S}{N} \rangle > 5$ region.

A1775: $\alpha/\beta$: excluded three point sources (12′ SE, 22′ SE, and 25′ S/SE). $\eta/\theta_o$: point source external to $\langle \frac{S}{N} \rangle > 5$ region. $w_{\bar{x}}$: $\langle \frac{S}{N} \rangle > 5$ region too small for measurement.

A1795: $\alpha/\beta$: excluded point source (5$'$ SW). $\eta/\theta_o/w_{\bar{x}}$: point source inhibits accurate measurements.
A1983: $w_{\bar{x}}$: $\left\langle\frac{S}{N}\right\rangle > 5$ region too small for accurate measurement.
A2029: $\alpha/\beta$: excluded two point sources (3$'$ SE and 15$'$ S/SW). $\eta/\theta_o/w_{\bar{x}}$: point source inhibits accurate measurements.
A2052: $\alpha/\beta$: excluded one point source (15$'$ NW). $\eta/\theta_o/w_{\bar{x}}$: point source is exterior to $\left\langle\frac{S}{N}\right\rangle > 5$ region.
A2063: $\alpha/\beta$: excluded one point source (3$'$ S/SE). $\eta/\theta_o/w_{\bar{x}}$: point source inhibits accurate measurement.
A2065: $\alpha/\beta$: excluded two point sources (10$'$ NW, 15$'$ W). $\eta/\theta_o/w_{\bar{x}}$: point sources are outside $\left\langle\frac{S}{N}\right\rangle > 5$ region.
A2124: $\alpha/\beta$: excluded possible point source (13$'$ NW). $\eta/\theta_o/w_{\bar{x}}$: point source outside $\left\langle\frac{S}{N}\right\rangle > 5$ region. $w_{\bar{x}}$: $\left\langle\frac{S}{N}\right\rangle > 5$ region too small for measurement.
A2142: $\alpha/\beta$: excluded two point sources (4$'$ NE, 17$'$ SE). $\eta/\theta_o/w_{\bar{x}}$: point source inhibits accurate measurement.
A2147: $\alpha/\beta$: excluded three possible point sources (20$'$ E/SE, 15$'$ S/SE, 20$'$ NW). No $\beta$ included because no fit with at least a 5% probability of being consistent with the data exists. $\eta/\theta_o/w_{\bar{x}}$: point sources lie outside $\left\langle\frac{S}{N}\right\rangle > 5$ region.
A2151: $\alpha/\beta$: system too complex for radial averaging. $\eta/\theta_o$: truncated at $R = 6.4'$ to exclude subclumps E of main clump. $w_{\bar{x}}$: included subclumps.
A2199: $\alpha/\beta$: excluded one point source (18$'$ NW). $\eta/\theta_o/w_{\bar{x}}$: point source external to $\left\langle\frac{S}{N}\right\rangle > 5$ region.
A2255: $\alpha/\beta$: excluded two point sources (13$'$ N, 20$'$ NW). $\eta/\theta_o/w_{\bar{x}}$: point source are outside $\left\langle\frac{S}{N}\right\rangle > 5$ region.
A2256: $\beta$: no $\beta$–fit with at least a 5% probability of being consistent with the data exists.
A2410: $\alpha/\beta$: too complex for radial averaging. $\eta$: complex nature made $\eta$ measurement impossible. $w_{\bar{x}}$: both clumps but no point sources included in $\left\langle\frac{S}{N}\right\rangle > 5$ region.
A2440: $\alpha/\beta$: system too complex for radial averaging. $\eta$: complex nature made $\eta$ measurement impossible. $w_{\bar{x}}$: $\left\langle\frac{S}{N}\right\rangle > 5$ region too small for measurement.
A2593: $\alpha/\beta$: excluded two point sources (7$'$ N, 15$'$ SW). $\eta/\theta_o$: point sources outside $\left\langle\frac{S}{N}\right\rangle > 5$ region. $w_{\bar{x}}$: $\left\langle\frac{S}{N}\right\rangle > 5$ region too small for measurement.
A2597: $w_{\bar{x}}$: $\left\langle\frac{S}{N}\right\rangle > 5$ region too small for measurement.
A2626: $\alpha/\beta$: excluded one point source (7$'$ NE). $w_{\bar{x}}$: $\left\langle\frac{S}{N}\right\rangle > 5$ region too small for measurement.
A2634: $\alpha/\beta$: excluded 2 point sources (15$'$ N/NW, 23$'$ S/SW) and subclump (12$'$ NW). Radial velocity and photometry study provides evidence that subclump is due to background cluster (Pinkney *et al.* 1993). $\eta/\theta_o/w_{\bar{x}}$: truncated at $R = 8.5'$ to exclude background cluster and point source.
A2657: $\alpha/\beta$: excluded two point sources (7$'$ W, 15$'$ SW). $\eta/\theta_o/w_{\bar{x}}$: truncated at $R = 6.4'$ to exclude point sources.
A2670: $\alpha/\beta$: excluded four point sources (10$'$ N, 15$'$ N, 15$'$ E/SE, 20$'$ E). $\eta/\theta_o$: point sources are outside $\left\langle\frac{S}{N}\right\rangle > 5$ region. $w_{\bar{x}}$: $\left\langle\frac{S}{N}\right\rangle > 5$ region too small for measurement.
A2877: $\alpha/\beta$: excluded point source (2$'$SW). $\eta/\theta_o/w_{\bar{x}}$: bright point source near center of cluster makes accurate measurements impossible.
A3158: $\alpha/\beta$: excluded two possible point sources (15$'$ S, 12$'$ E/SE). $\eta/\theta_o/w_{\bar{x}}$: point sources are outside $\left\langle\frac{S}{N}\right\rangle > 5$ region.
A3186: $w_{\bar{x}}$: $\left\langle\frac{S}{N}\right\rangle > 5$ too small for measurement.
A3266: $\alpha/\beta$: large core radius contributes to large $\beta$.
A3376: $\alpha/\beta$: excluded one point sources (20$'$ S/SW). $\eta/\theta_o/w_{\bar{x}}$: no rib or point source contamination within $\left\langle\frac{S}{N}\right\rangle > 5$ region.
A3391: $\alpha/\beta$: excluded one point source (13$'$ N/NE). $\eta/\theta_o/w_{\bar{x}}$: point source outside $\left\langle\frac{S}{N}\right\rangle > 5$ region.
A3395: $\alpha/\beta$: bimodal structure makes radial averaging pointless. $\eta/\theta_o$: considered each subclump (NE and SW) separately. $w_{\bar{x}}$: Peak of the brighter (SW) subclump used as the true cluster center.
A3526: $\eta/\theta_o/w_{\bar{x}}$: truncated at $R = 9.6$ to avoid rib contamination.
A3532: $\alpha/\beta$: excluded extended emission near ribs to W. $w_{\bar{x}}$: $\left\langle\frac{S}{N}\right\rangle > 5$ region too small for measurement.
A3667: $\alpha/\beta$: excluded three point sources (12$'$ NW, 20$'$ E/SE, 15$'$ S/SW). $\eta/\theta_o/w_{\bar{x}}$: truncated at $R = 7.5'$ to avoid point source contamination.
CygA: $\alpha/\beta$: excluded subclump 12$'$ NW. $\eta/\theta_o$: truncated at $R = 8.5'$ to avoid subclump. $w_{\bar{x}}$: subclump included.
MKW3S: $w_{\bar{x}}$: $\left\langle\frac{S}{N}\right\rangle > 5$ region too small for measurement.
3C129: $\alpha/\beta$: excluded one point source (15$'$ N). No $\beta$ because the best fit core radius is comparable to the IPC field of view. $\eta/\theta_o$: truncated at $R = 11.7'$ to avoid point source contamination.